\documentclass[letterpaper]{JHEP3}
\usepackage{array}
\usepackage{nicefrac}
\usepackage{amsmath,epsfig}
\usepackage{amssymb,amsfonts,mathrsfs}

\usepackage{graphics}
\usepackage{epsfig}
\usepackage{amsfonts}

\newcommand{\Tr}{\mbox{Tr}}

\newcommand{\rep}[1]{\mathbf{#1}}
\newcommand{\crep}[1]{\overline{\mathbf{#1}}}
\newcommand{\sym}[2]{sym^{#1}(#2)}

\global \long \def \NN{ \mathcal{N}}

\def\e{\epsilon}

\def\h{\eta}

\def\IC{\relax\hbox{$\inbar\kern-.3em{\rm C}$}}

\def\IC{{\bf C}}

\def\bea{\begin{eqnarray}}
\def\eea{\end{eqnarray}}
\def\be{\begin{equation}}
\def\ee{\end{equation}}
\def\ba{\begin{align}}
\def\ea{\end{align}}
\def\bse{\begin{subequations}}
\def\ese{\end{subequations}}
\def\1F1{{}_1\!F_1}
\def\2F0{{}_2\!F_0}

\def\h3{$\textrm{H}_3^+$}

\def\IC{{\mathbb C}}

\def\tr{{\rm tr}}
\def\Tr{{\rm Tr}}

\def\lbldef#1#2{\expandafter\gdef\csname #1\endcsname {#2}}

\def\href#1#2{#2}

\newcommand{\beq}{\begin{equation}}
\newcommand{\eeq}{\end{equation}}
\newcommand{\ber}{\begin{eqnarray}}
\newcommand{\eer}{\end{eqnarray}}

\def\be{\begin{eqnarray}}
\def\ee{\end{eqnarray}}

\providecommand{\tabularnewline}{\\}

\def\({\left(}
\def\){\right)}
\def\[{\left[}
\def\]{\right]}
\def\<{\langle}
\def\>{\rangle}

\def\tr{{\rm Tr}}

\title{The Superconformal Index of the $E_6$ SCFT}
\preprint{YITP-SB-10-7}
\author{Abhijit Gadde\footnote{abhijit@insti.physics.sunysb.edu},
 Leonardo Rastelli\footnote{leonardo.rastelli@stonybrook.edu},
Shlomo S. Razamat\footnote{razamat@max2.physics.sunysb.edu},
and Wenbin Yan\footnote{wyan@insti.physics.sunysb.edu}
\\
\\
\it C.N. Yang Institute for Theoretical Physics,\\
\it Stony Brook University, \\
\it Stony Brook, NY 11794-3840, USA}

\bigskip

\abstract{ We  derive an  integral representation for the superconformal
index of the strongly-coupled $\NN=2$ superconformal field theory 
with $E_6$ flavor symmetry. The explicit expression of the index 
allows highly non-trivial checks of
 Argyres-Seiberg duality and of a class of S-dualities
conjectured  by Gaiotto. }


\begin{document}
\section{Introduction}

The paradigmatic S-duality of ${\cal N}=4$ super Yang-Mills
is the simplest instance of a much more general web of duality connections
relating ${\cal N}=2$ $4d$ superconformal field theories. This viewpoint has been
emphasized by Gaiotto~\cite{Gaiotto:2009we}, who introduced a large class of ${\cal N} =2$
SCFTs by compactifying the $(2,0)$ $6d$ theory on a Riemann
surfaces $\Sigma$ with punctures. Different ways of cutting $\Sigma$
into  pairs of pants correspond to different S-duality frames for the $4d$ theory.
A remarkable dictionary relates  $4d$ gauge theory quantities with calculations in
 $2d$ conformal field theory on $\Sigma$.
For example, the partition function of the gauge theory on $S^4$, or more generally
the Nekrasov
instanton partition function~\cite{Nekrasov:2002qd}, is 
 reproduced exactly  by a Liouville or Toda correlation function  on $\Sigma$~\cite{Alday:2009aq,
Wyllard:2009hg}.

This dictionary was extended in~\cite{Gadde:2009kb}
by considering the superconformal index~\cite{Kinney:2005ej}, 
which can be viewed as a twisted partition function of the $4d$ gauge theory $S^3 \times S^1$.
The superconformal index   counts the states of the $4d$ theory
 belonging to short multiplets, up to  equivalent relations that set to 
 to zero all sequences of short multiplets that may in principle recombine into long ones.
By construction, the index is invariant under continuous deformations of the theory, and is also expected to be independent
of the  S-duality frame. {\it Assuming} S-duality, it follows that the index must be computed by a topological QFT living on $\Sigma$.
In~\cite{Gadde:2009kb} this TQFT structure was discussed for the generalized quiver gauge theories with  $SU(2)^{k}$ gauge group,
which arise from compactifications on $\Sigma$  of the $A_1$ (2,0) theory.  Invariance of the index under
S-duality translates into associativity of the operator algebra of the $2d$ TQFT. In turn, associativity
holds thanks to a beautiful mathematical identity for an elliptic hypergeometric integral~\cite{Focco}.

What distinguishes the $A_1$ theories from their counterparts with $A_{n \geq 2}$ 
is that in all duality frames they have a  Lagrangian description.
This makes it easy to compute their superconformal index explicitly and to identify the structure
constants of the $2d$ TQFT \cite{Gadde:2009kb}.
The situation for the generalized quiver theories with higher rank gauge groups is qualitatively different:
in some  duality frames the quivers contain intrinsically strongly-coupled blocks with no Lagrangian
description. The prototypical example of this phenomenon was discussed by  Argyres and Seiberg \cite{argyres-2007-0712}\footnote{See also~\cite{Argyres:2007tq}
for more examples.}:
  the  SYM theory with  $SU(3)$ gauge group and $N_f =6$ fundamental hypermultiplets has a dual description involving
 the strongly-coupled  SCFT with $E_6$ flavor symmetry~\cite{Minahan:1996fg}. In the absence of a Lagrangian
 description for the $E_6$ SCFT, it seems difficult to compute its superconformal index and to
define the TQFT structure for generalized quivers with $SU(3)$ gauge groups.

We solve this problem in this paper. By demanding consistency with Argyres-Seiberg duality, we are
able to write down an explicit integral expression for the index of the $E_6$ SCFT (equation~\eqref{strongIndex}).
Technically, this is possible thanks to a remarkable inversion formula  for  a class of integral transforms~\cite{spirinv}. 
By construction, the resulting expression for the  index is guaranteed to be invariant under an $SU(6) \otimes SU(2)$ subgroup
of the $E_6$ flavor symmetry. The index is seen {\it a posteriori} to be invariant under the full $E_6$ symmetry, providing an
independent check of Argyres-Seiberg duality itself.\footnote{For earlier checks of Argyres-Seiberg duality see~\cite{Aharony:2007dj}
 and~\cite{Gaiotto:2008nz}.} We proceed 
 to define a TQFT structure for generalized quivers with $SU(3)$ gauge symmetries. We check associativity
 of the operator algebra, which is equivalent to a check of S-duality for Gaiotto's $A_2$  theories.
Most of our checks are performed perturbatively, to  several orders in an expansion in  the chemical potentials
that enter the definition of the index. Conversely, S-duality implies 
that associativity must hold exactly, so as a by-product of our analysis we conjecture
new  identities between integrals of elliptic Gamma functions.

The paper is organized as follows. In section \ref{genSec} we set up the stage by
 briefly reviewing  the definitions of the superconformal index
and of the elliptic Gamma functions. In section \ref{weakSec} the index of $N_f=6$ $SU(3)$ theory
 is computed in the weakly-coupled frame
and the usual S-duality invariance of this index is discussed. In section \ref{strongSec} we use  Argyres-Seiberg duality
to write down an explicit expression for the index of $E_6$ SCFT; we check perturbatively that the answer is $E_6$ covariant and that it
is compatible with physical expectations about the Coulomb and Higgs branches of vacua.
  In section \ref{sdualSec} 
we check  invariance under S-duality of the superconformal index for the generalized $SU(3)$ quiver theories,
 and we present the TQFT interpretation of this index. In section \ref{discSec} we briefly discuss our results.
Four appendices complement the text with technical details.

\section{Generalities}\label{genSec}

In this section we briefly review the definition of the superconformal index~\cite{Kinney:2005ej}, and the relevant properties of elliptic Gamma functions.
\subsection{The superconformal index}
The superconformal index is defined as~\cite{Kinney:2005ej}\footnote{See also~\cite{Romelsberger:2005eg}.}
 \be
 \mathcal{I}=  \mbox{Tr}(-1)^{F}t^{2(E+  j_2)}y^{2\,j_1}v^{-(r+R)} \,,
 \ee
 where we trace  over the states of the theory on $S^3$ (in the usual radial quantization).\footnote{
For definiteness we  consider the ``right-handed'' Witten index ${\cal I}^{WR}$ of \cite{Kinney:2005ej}, which computes
 the cohomology of the supercharge $\bar {\cal Q}_{2 +}$. We use the notations of \cite{Dolan:2002zh}
 where the supercharges are denoted
 as ${\cal Q}^I_\alpha$, $\bar {\cal Q}_{I  \dot \alpha}$, ${\cal S}_{I \alpha}$, $\bar {\cal S}^{I}_  {\dot \alpha}$,
 with $I =1,2$ $SU(2)_R$ indices and $\alpha=\pm$, $\dot \alpha = \pm$ Lorentz indices.} The chemical potentials $t$, $y$, and $v$
 keep track of various combinations of quantum numbers associated to the supercorformal algebra $SU(2,2 |2)$:
 $E$ is the conformal dimension, $(j_1, j_2)$ the $SU(2)_1 \otimes SU(2)_2$  Lorentz spins, and $(R \, ,r)$ the quantum numbers
 under the  $SU(2)_R \otimes U(1)_r$ R-symmetry.\footnote{Our normalization convention for the R-symmetry charges is as in~\cite{Dolan:2002zh}
and  differs from \cite{Kinney:2005ej}: $R_{here} = R_{there}/2$, $r_{here} = r_{there}/2$.}

 For a theory with a weakly-coupled description the index can be explicitly computed as a matrix integral,
\be
\label{index}
{\cal I}(V,t,y,v) =\int\left[dU\right]\, \exp\left(\sum_{n=1}^\infty\frac{1}{n}\;\sum_{j} f^{\mathcal R_j}(t^n,y^n,v^n)  \cdot \chi_{{\mathcal R_j}}(U^n,\,V^n)\right) \, .
\ee Here $U$ is the matrix of the gauge group,  $V$  the matrix of the flavor group and ${\mathcal R_j}$ label representations of the fields
under the flavor and  gauge groups. The measure $\left[dU\right]$ is the invariant Haar measure, and it has the following property
\be
\int\left[dU\right]\,\prod_{j=1}^n \chi_{{\mathcal R_j}}(U)=\# {\text{of singlets in}\; } {\mathcal R_1}\otimes\dots\otimes {\mathcal R_n}\,.
\ee
\begin{table}
  \begin{centering}
  \begin{tabular}{|c|r|r|r|r|r|c|}
  \hline
Letters & $  E$ & $j_1$ & $  j_2$ & $R$ & $r$ & $  \mathcal{I}$  \tabularnewline
  \hline
   \hline
$  \phi$ & $1$ & $0$ & $0$ & $0$ & $-1$ & $t^{2}v$  \tabularnewline
  \hline
$  \lambda_{\pm}^{1}$ & $  \frac{3}{2}$ & $  \pm  \frac{1}{2}$ & $0$ & $  \frac{1}{2}$ & $-  \frac{1}{2}$ & $-t^{3}\,y,\;-t^3\,y^{-1}$  \tabularnewline
  \hline
$  \bar{\lambda}_{2+}$  & $  \frac{3}{2}$ & $0$ & $  \frac{1}{2}$ & $  \frac{1}{2}$ & $  \frac{1}{2}$ & $-t^{4}/v$  \tabularnewline
  \hline
$  \bar{F}_{++}$ & $2$ & $0$ & $1$ & $0$ & $0$ & $t^{6}$  \tabularnewline
  \hline
  $  \partial_{-+}  \lambda_{+}^{1}+  \partial_{++}  \lambda_{-}^{1}=0$ & $  \frac{5}{2}$ & $0$ & $  \frac{1}{2}$ & $  \frac{1}{2}$ & $  -\frac{1}{2}$ & $t^{6}$ \tabularnewline
  \hline
\hline
$q$ & $1$ & $0$ & $0$ & $  \frac{1}{2}$ & $0$ & $t^{2}/  \sqrt{v}$  \tabularnewline
  \hline
$  \bar{\psi}_{+}$ & $  \frac{3}{2}$ & $0$ & $  \frac{1}{2}$ & $0$ & $-  \frac{1}{2}$ & $-t^{4}  \sqrt{v}$  \tabularnewline
  \hline
    \hline
$  \partial_{\pm+}$ & $1$ & $  \pm  \frac{1}{2}$ & $  \frac{1}{2}$ & $0$ & $0$ & $t^{3\,}y,\;t^3\,y^{-1}$  \tabularnewline
  \hline
  \end{tabular}
  \par  \end{centering}
  \caption{Contributions to the index from  ``single letters''.
  We denote by $(\phi, \bar \phi,  \lambda^I_\alpha, \lambda_{I \,\dot \alpha},  F_{\alpha \beta}, \bar F_{\dot \alpha \dot \beta})$
 the components of the adjoint ${\cal N} = 2$ vector multiplet,  by $(q, \bar q, \psi_\alpha, \bar \psi_{\dot \alpha})$ the components  of the  ${\cal N} = 1$
chiral multiplet,  and by $\partial_{\alpha \dot \alpha}$ the spacetime derivatives. Here $I = 1,2$ are $SU(2)_R$ indices and
$\alpha = \pm$, $\dot \alpha = \pm$ Lorentz indices.
}
\label{letters}
  \end{table}
The quantities $f^{\mathcal R_j}(t,y,v)$ are the single-letter partition functions for matter in representation ${\mathcal R_j}$.
The ``single letters'' of an ${\mathcal N}=2$ gauge theory contributing to the index must obey $
 E - 2 j_2 - 2 R +r = 0$  \cite{Kinney:2005ej} and are enumerated in table \ref{letters}.
The first block of table~\ref{letters} shows the contributing letters from the ${\cal N} = 2$ vector multiplet,
including the equations of motion constraint. The second block shows the contributions from the half hypermultiplet (or ${\cal N}= 1$ chiral multiplet). The
   last line shows the spacetime derivatives
contributing to the index. Since each field can be hit by an arbitrary number of derivatives, the derivatives give
a multiplicative contribution to the single-letter partition functions of the form
\be
\sum_{m=0}^\infty \sum_{n=0}^\infty (t^3 y)^m  \, (t^3 y^{-1})^n = \frac{1}{(1-t^3 y)(1-t^3 y^{-1})}\,.
\ee
The single-letter partition functions of the ${\cal N}=2$ vector and ${\cal N}=1$ chiral multiplets are thus given by
\be\label{letterpart}
\text{vector}\qquad&:&\qquad  f^{vect}(t,y,v)=\frac{t^2v-\frac{t^4}{v}-t^3(y+y^{-1})+2t^6}{(1-t^3\,y)(1-t^3y^{-1})}\,,\\
{\rm chiral} \qquad&:&\qquad  f^{chi}(t,y,v)=\frac{\frac{t^2}{\sqrt{v}}-t^4\sqrt{v}}{(1-t^3\,y)(1-t^3y^{-1})}\,.
\ee
Throughout this paper we will assume
\be
0<|t|^4<|v|<1\,. 
\ee

\subsection{Elliptic hypergeometric expressions for the index}
As was observed by Dolan and Osborn~\cite{Dolan:2008qi} the expressions for the index can be recast in an elegant way
in terms of special functions.
First, recall the definition of  the elliptic Gamma function,
\be\label{gamma1}
\Gamma(z;p,q) \equiv \prod_{j,k\geq 0}\frac{1-z^{-1}\,p^{j+1}q^{k+1}}{1-z\,p^jq^k} \,.
\ee
For reviews of the elliptic Gamma function and of elliptic
hypergeometric mathematics the reader can consult~\cite{Spiridonov,Spiridonov2,Spiridonov3}.
 Throughout this paper we will use the standard condensed notations
\be\label{condensed}
&&\Gamma(z_1,\dots, z_k;p,q)\equiv \prod_{j=1}^k \Gamma(z_j;p,q),\qquad \Gamma(z^{\pm1};p,q)\equiv \Gamma(z;p,q)\Gamma(1/z;p,q) \,.
\ee
Basic identities satisfied by the elliptic Gamma function that will be of use to us are
\be
&&\Gamma\left(pq/z;p,q\right)\Gamma\left(z;p,q\right)=1\,,\\\label{inverse}
&& \lim_{z\rightarrow a}\(1-z/a\)\Gamma(z/a;p,q)=\frac{1}{(p;p)(q;q)}\,,\label{limGamm}
\ee with the bracket defined as
\be
\left(a;\, b\right) \equiv \prod_{k=0}^\infty \left(1-a\,b^k\right)\,.
\ee
From the definition (\ref{gamma1}),
it is  straightforward to show~\cite{Dolan:2008qi}
\be\label{dolanid}
&&\exp\left({\sum_{n=1}^\infty \frac{1}{n}}\,\frac{t^{2n}z^n-t^{4n}z^{-n}}{(1-t^{3n} y^n)(1-t^{3n}y^{-n})}\right)=\Gamma(t^2\,z; p,q),\\
&&\exp\left({\sum_{n=1}^\infty \frac{1}{n}}\,\frac{2t^{6n}-t^{3n}(y^n+y^{-n})}{(1-t^{3n} y^n)(1-t^{3n}y^{-n})}(z^n+z^{-n})\right)=
-\frac{z}{(1-z)^2}\,\frac{1}{\Gamma(z^{\pm1};p,q)},\nonumber
\ee
where
\be
p=t^3y,\qquad q=t^3y^{-1} \, .
\ee Using the above identities the basic building blocks of the superconformal
index computation can be written as follows. The contribution to the integrand
of \eqref{index} from hypers in a fundamental representation of an $SU(n)$ gauge group is
\be\label{hyper}
\exp\left(\sum_{k=1}^\infty\frac{1}{k}\,f^{chi}\left(t^k,\,v^k,\, y^k\right)\,\left[\chi_{f}(U^k)+\chi_{\bar f}(U^k)\right]\right)= \prod_{i=1}^n\,\Gamma\left(\frac{t^2}{\sqrt{v}}a_i^{\pm1};p,q\right)\,.\nonumber\\
\ee
The contribution to the integrand of \eqref{index} from the vector multiplet of $SU(n)$ is
\be
&&\exp\left(\sum_{k=1}^\infty\frac{1}{k}\,f^{vect}\left(t^k,\,v^k,\, y^k\right)\;\chi_{adj}(U^k)\;\right)=\\%\nonumber\\
&&\qquad\qquad\qquad
= \frac{\left[\Gamma(t^2\, v;p,q)\,(p;p)(q;q)\right]^{n-1}}{\Delta(\mathbf{a})\Delta(\mathbf{a}^{-1})}\,\,\prod_{i\neq j}\frac{\Gamma(t^2\, v\,a_i/a_j;p,q)}{\Gamma(a_i/a_j;p,q)}\, .\nonumber
\ee We have defined the characters of the fundamental representation to be
\be
\chi_{f}=\sum_{i=1}^n a_i,\qquad \chi_{\bar f}=\sum_{i=1}^n \frac1a_i,\qquad \prod_{i=1}^n a_i=1\,.
\ee The character of the adjoint representation is
\be
\chi_{adj}=\chi_{f}\, \chi_{\bar f}-1=\sum_{i\neq j} a_i/a_j +n-1\,.
\ee
We have also defined
\be
\Delta(\mathbf{a})=\prod_{i\neq j} (a_i-a_j)\,.
\ee
The Haar measure is given by
\be
\oint_{SU(n)}d\mu({\mathbf a})f({\mathbf a})=
\frac{1}{n!}\oint_{\mathbb{T}^{n-1}}\left.\prod_{i=1}^{n-1}\frac{da_{i}}{2\pi i\,a_{i}}\Delta({\mathbf{a}})\Delta({\mathbf{a}}^{-1})
f({\mathbf a})\right|_{\prod_{i=1}^{n}a_{i}=1}\,,
\ee where $\mathbb{T}$ is the unit circle. Whenever we gauge a symmetry we have a vector multiplet associated to the integrated group and thus we will use the following
notation
\be
{\mathcal F}_{\mathbf a}\,{\mathcal G}^{\mathbf a}\equiv
\frac{\left[2\,\Gamma(t^2\, v;p,q)\,\kappa\right]^{n-1}}{n!}\oint_{\mathbb{T}_{n-1}}\left.\prod_{i=1}^{n-1}\frac{da_{i}}{2\pi i\,a_{i}}\,\prod_{i\neq j}\frac{\Gamma(t^2\, v\,a_i/a_j;p,q)}{\Gamma(a_i/a_j;p,q)}
{\mathcal F}\left({\mathbf a}\right)\,{\mathcal G}\left({\mathbf a}^{-1}\right)\right|_{\prod_{i=1}^{n}a_{i}=1}\, ,\nonumber\\
\ee where $\kappa\equiv (p;p)(q;q)/2$.
 In what follows for the sake of brevity we will omit the parameters $p$ and $q$ from the elliptic Gamma function, i.e.
 $\Gamma(x)$ should always be understood as $\Gamma(x;\,p,q)$.

\section{Argyres-Seiberg duality and the index of $E_6$ SCFT}
\begin{figure}
\begin{center}
$\begin{array}{c@{\hspace{0.5in}}c@{\hspace{0.0in}}c}
\epsfig{file=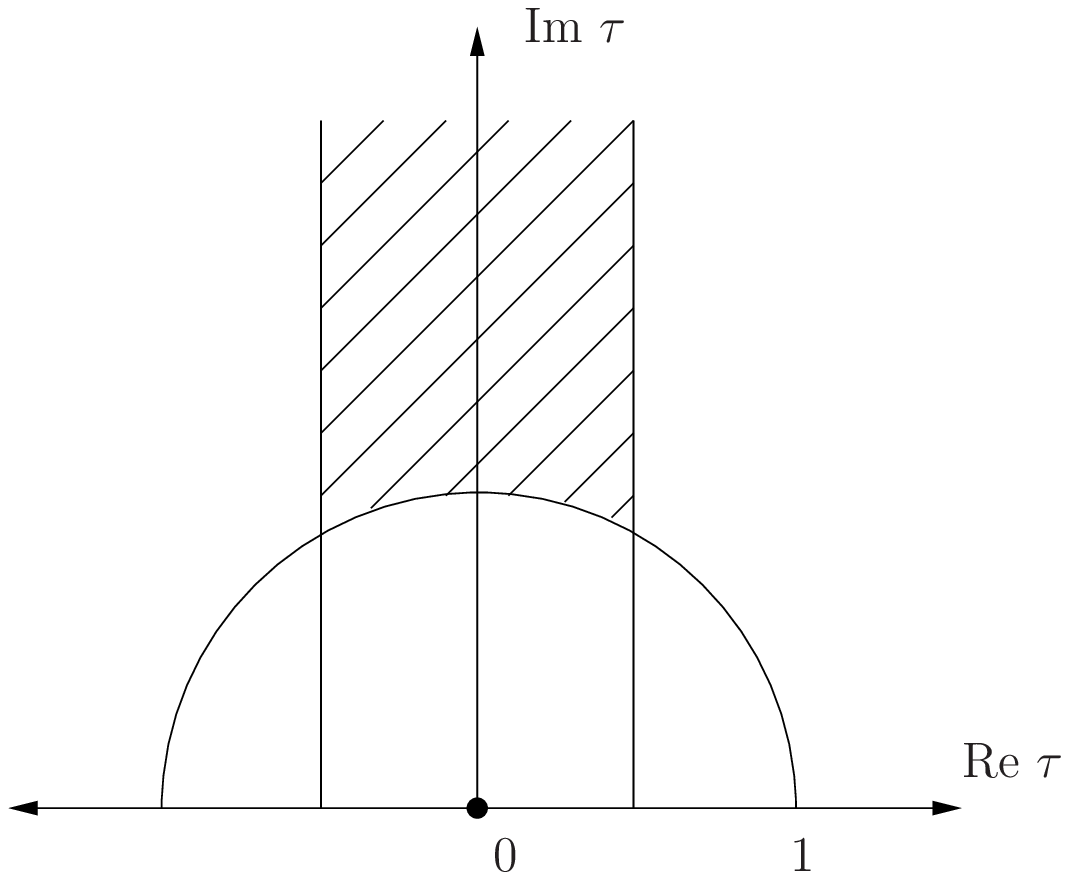,scale=0.6} & \epsfig{file=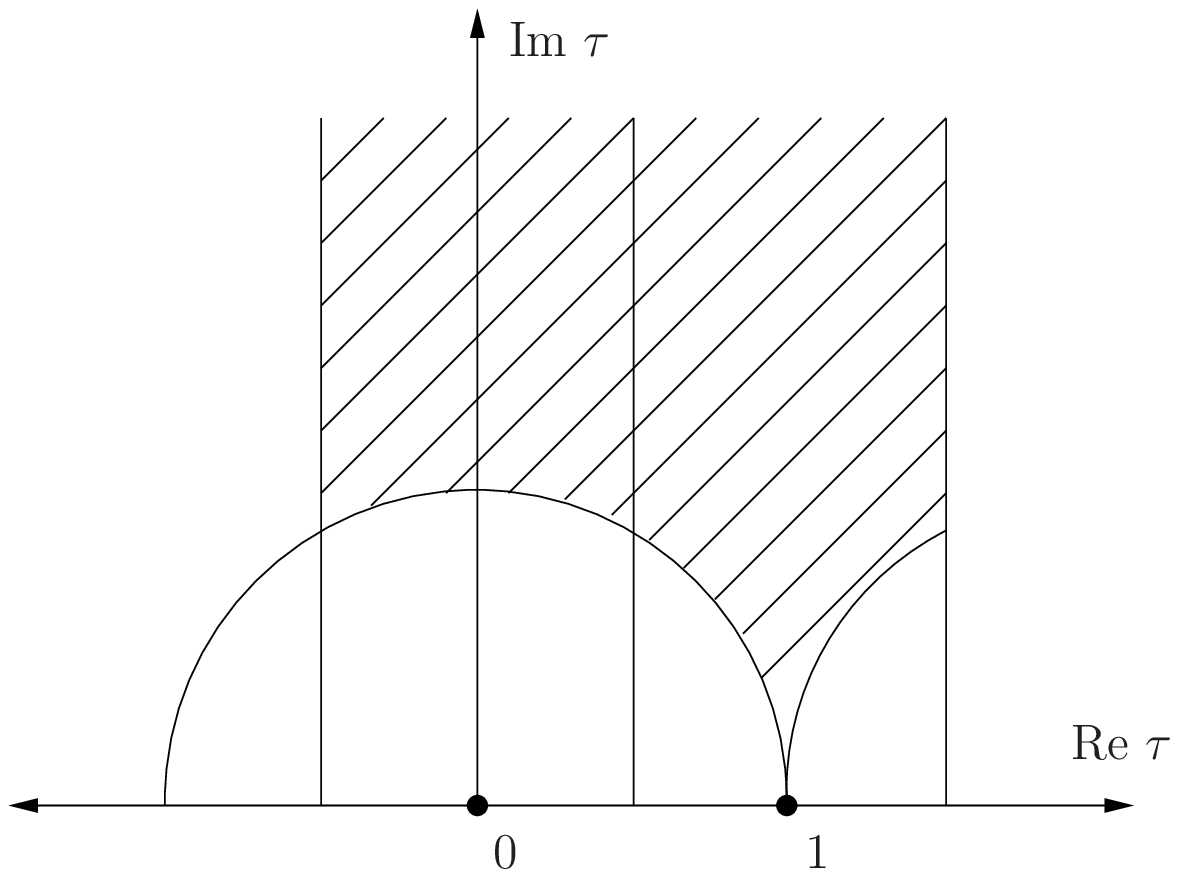,scale=0.6}\\\
(a)& (b)
\end{array}$
\end{center}
 \begin{center}
\caption
{Moduli spaces for $\NN=2$ $SU(n)$ gauge theory with $2n$ flavors, (a) for $n=2$ and (b) for $n=3$ (in fact, for any $n >2$). The shaded region in (a) is $H/SL(2,\mathbb Z)$ while in (b) it is $H/\Gamma^0(2)$, where $H$ is the upper half plane.
} \label{modulispace}
\end{center}
\end{figure}

The S-duality group of the $\NN=2$ $SU(2)$ gauge theory with four flavors is $SL(2,\mathbb {Z})$. The action of this group on the gauge coupling is generated by $\tau\to\tau+1$ and $\tau\to -1/\tau$. In Gaiotto's description~\cite{Gaiotto:2009we} this theory is constructed by compactification of the $6d$ $(2,0)$ theory on a sphere with four punctures of the \emph{same} kind. Then, the S-duality group could be understood as the mapping class group of this Riemann surface.
The moduli space of the gauge coupling is shown in figure \ref{modulispace} (a). We can see that a fundamental domain can be chosen such that nowhere in the moduli space does the coupling take an infinite value.

For the case of $\NN=2$ $SU(3)$ gauge theory with $6$ flavors,
however, the S-duality group is $\Gamma^0(2)$. The action of the S-duality on the
complex coupling is generated by the transformations $\tau\to\tau+2$
and $\tau\to-1/\tau$. In Gaiotto's setup this theory is obtained by
compactifying the $(2,0)$ theory on the sphere with two punctures of
one type and two of another. The mapping class group of such a
sphere is $\Gamma^0(2)$. The fundamental domain of this group is
shown in the figure \ref{modulispace} (b) and, unlike the $SU(2)$
case, this does unavoidably contain a point with infinite coupling. In
\cite{argyres-2007-0712}, it was shown that this infinitely coupled
cusp could be described in terms of an $SU(2)$ gauge group weakly-coupled to a single hypermultiplet and a rank $1$ interacting SCFT
with $E_6$ flavor symmetry. Figure \ref{strongFig} describes this
duality pictorially. The $SU(2)$ subgroup of the flavor symmetry of
the SCFT that is gauged commutes with the $SU(6)$ subgroup of $E_6$.
This $SU(6)$ combined with $SO(2)$ flavor symmetry of the single
hypermultiplet generates the full $U(6)$ flavor symmetry of the
original $SU(3)$ gauge theory. In other words, the $SO(2)$ flavor
symmetry of the single hypermultiplet corresponds to the baryon
number of the original $SU(3)$ gauge theory. The quarks of the
$SU(3)$ theory are charged $\pm1$ under this $U(1)_B$ while the
quarks of the $SU(2)$ theory are charged $\pm3$ under the same.

The $E_6$ SCFT has a Coulomb branch parametrized by the expectation
value of a dimension $3$ operator $u$ which is identified with $\Tr
\phi^3$ of the dual $SU(3)$ theory, while the $\Tr\phi^2$ of the
$SU(3)$ theory corresponds to the Coulomb branch parameter of the
$SU(2)$ gauge theory. The $E_6$ CFT also has a Higgs branch
parametrized by the expectation value of dimension $2$ operators
$\mathbb{X}$, which transform in the adjoint representation of $E_6$
($\rep{78}$). As shown in~\cite{Gaiotto:2008nz} the Higgs branch
operators obey a Joseph relation at quadratic order which leaves a
$22$ complex dimensional Higgs branch. When coupled to the $SU(2)$
gauge group, the resulting Higgs branch has complex dimension $20$.
The dual $SU(3)$ theory also has a Higgs branch of complex dimension
$20$ and its Higgs operators can be easily constructed by
combination of squark fields.
%Coulomb and Higgs branch operators could be identified with their counterparts on the other side of the duality using the quantum numbers of the operators in the $E_6$ SCFT. 
See appendix~\ref{CoulHiggs} for more details.

\begin{figure}
\begin{centering}
\includegraphics[scale=0.7]{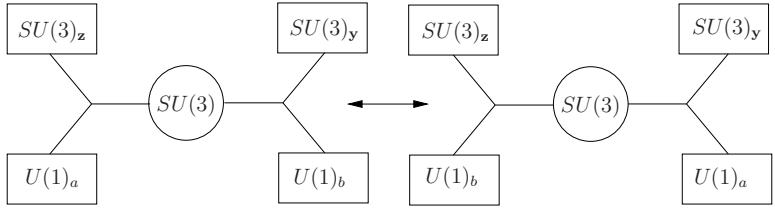}
\par\end{centering}

\caption{\label{duality1}
$SU(3)$ SYM with $N_f = 6$. The $U(6)$ flavor symmetry is decomposed  as 
$SU(3)_{\mathbf z} \otimes U(1)_a \oplus SU(3)_{\mathbf y} \otimes U(1)_b$.   S-duality $\tau \to -1/\tau $ interchanges the two $U(1)$ charges.}
\end{figure}

The moduli space might contain also other infinitely coupled cusps which however
are S-dual to the weakly-coupled cusp $\tau=i \infty$. This is the usual S-dualty mapping the $N_f=6$
$SU(3)$ gauge theory to itself with some of the $U(1)$ flavor factors
 interchanged. This duality is represented in figure
\ref{duality1}.

We proceed to compute the superconformal index of the $SU(3)$
 theory and, by using the Argyres-Seiberg duality,
of the interacting $E_6$ SCFT.

\subsection{Weakly-coupled frame}\label{weakSec}
\label{sec:IndexinArgyres-SeibergDuality}

We take the chiral multiplets to be in the fundamental and
antifundamental of the color and flavor. $U(1)_{B}$ rotates them
into each other. The vector multiplet is in the adjoint of
the color. The $SU(3)$ characters of the relevant representations are:
\be
\chi_{f}=z_{1}+z_{2}+z_{3}\quad\chi_{\bar{f}}=\frac{1}{z_{1}}+\frac{1}{z_{2}}+\frac{1}{z_{3}}\quad\mbox{and}\quad\chi_{adj}=\chi_{f}\chi_{\bar{f}}-1
\ee
while writing down these characters, we have to impose
$z_{1}z_{2}z_{3}=1$.

Let $z$'s stand for the eigenvalues of the flavor group and $x$'s be
the eigenvalues of the color group. The $U(1)_{B}$ charge is counted
by the variable $a$. Let us write down the characters of the
representation of the matter
\begin{equation}
  \label{eq:character:SU(3)}
    \chi_{hyp}
             =  \sum_{i=1}^{3}\sum_{j=1}^{3}a\,z_{i}\,x_{j}
                 +\sum_{i=1}^{3}\sum_{j=1}^{3}\frac{1}{a\,z_{i}\,x_{j}}\,.
\end{equation}
 Using~\eqref{hyper} the index contributed by the matter
can be written in a closed form as
\be\label{vert331}
C_{a,{\mathbf x},{\mathbf y}}=\prod_{i=1}^3\prod_{j=1}^3\, \Gamma\left(\frac{t^2}{\sqrt{v}}\,\left(a \, x_i\,y_j\right)^{\pm1}\right) \,.
\ee
The index for the $SU(3)$ gauge theory with six hypermultiplets is
then given by the following contour integral.
\begin{eqnarray}\label{indexWeak}
&&\mathcal{I}_{a,{\mathbf z};b,{\mathbf y}}  =  C_{b,{\mathbf y},{\mathbf x}}\,{C_{a,{\mathbf z}}}^{\mathbf x}=\\
&&\qquad\frac{2}{3}\kappa^2\Gamma(t^{2}v)^{2}
 \oint_{\mathbb{T}^{2}}\prod_{i=1}^{2}\frac{dx_{i}}{2\pi i\,x_{i}}
\frac{\displaystyle\prod_{i=1}^3\prod_{j=1}^3\Gamma\left(\frac{t^{2}}{\sqrt{v}}\left(\frac{az_{i}}{x_{j}}\right)^{\pm1}\right)
\Gamma\left(\frac{t^{2}}{\sqrt{v}}\left(b\,y_{i}\,x_{j}\right)^{\pm1}\right)\prod_{i\neq j}
\Gamma\left(t^{2}v\,\frac{x_{i}}{x_{j}}\right)}{\displaystyle\prod_{i\neq j}\Gamma\left(\frac{x_{i}}{x_{j}}\right)}.\nonumber
\end{eqnarray}
By expanding this integral in $t$ one can show that it is symmetric under interchanging the two $U(1)$ factors (see appendix~\ref{LHS}),
\be\label{weakSymm}
a\quad\leftrightarrow \quad b\,.
\ee
Interchanging the two $U(1)$s is equivalent to performing a usual S-duality between a weakly-coupled
and infinitely-coupled points of the  moduli space
and thus we expect the index to be invariant under this operation.\footnote{The integral~\eqref{indexWeak} is
an $SU(3)$ generalization of the $SU(2)$ integral in~\cite{Gadde:2009kb} for which the analogous statement to~\eqref{weakSymm}
has an analytic proof~\cite{Focco}. It is easy to generalize~[\ref{indexWeak},\ref{weakSymm}] for $SU(n)$ theories with arbitrary $n$,
see appendix~\ref{identApp}.}

One can analytically prove this statement in a special case.
Notice that if $t=v$, the integral~\eqref{indexWeak}  is given by
\be
 \left.\mathcal{I}_{a,{\mathbf z};b,{\mathbf y}}\right|_{v=t}=
I_{A_{2}}^{(2)}\left(\left.1\right|t^{\frac{3}{2}}a^{-1}{\mathbf
z}^{-1},t^{\frac{3}{2}}b{\mathbf y};t^{\frac{3}{2}}a{\mathbf
z},t^{\frac{3}{2}}b^{-1}{\mathbf y}^{-1}\right)\, ,
\ee  where~\cite{rains}
\be
&&I_{A_{n}}^{(m)}(Z|t_0,\dots,t_{n+m+1};u_0,\dots,u_{n+m+1};\,p,q)=\\
&&\qquad\qquad\qquad\frac{2^n}{n!}\kappa^n\oint_{\mathbb{T}^{n-1}}\left.\prod_{i=1}^{n-1}\frac{dx_i}{2\pi
i\, x_i}\,
\frac{\prod_{i=1}^n\prod_{j=0}^{m+n+1}\Gamma(t_j\,x_i,\,u_j/x_i;\,p,q)}{\prod_{i\neq
j}\Gamma(x_i/x_j;\,p,q)}\right|_{\prod_{i=1}^nx_i=Z}\,.\nonumber \ee
If the integral $I_{A_{n}}^{(m)}(Z|\ldots t_{i}\ldots;\ldots
u_{i}\ldots)$ satisfies the condition that
$\prod_{i=1}^{m+n+2}t_{i}u_{i}=(pq)^{m+1}$ then due to \cite{rains},
the following theorem holds \be\label{Rains}
I_{A_{n}}^{(m)}\left(Z|\ldots t_{i}\ldots;\ldots
u_{i}\ldots\right)=I_{A_{m}}^{(n)}\left(Z|\ldots\frac{T^{\frac{1}{m+1}}}{t_{i}}\ldots;\ldots\frac{U^{\frac{1}{m+1}}}{u_{i}}\ldots\right)\prod_{r,s=1}^{m+n+2}\Gamma\left(t_{r}u_{s}\right)\,,
\ee where $T\equiv\prod_{r=1}^{m+n+2}t_{r}$ and
$U\equiv\prod_{r=1}^{m+n+2}u_{r}$.\footnote{This identity was
extensively used in~\cite{Dolan:2008qi} to show that certain
theories related by Seiberg duality have equal superconformal
indices~\cite{Romelsberger:2007ec}. In this context the authors
of~\cite{Spiridonov:2008zr,Spiridonov:2009za} applied the elliptic
hypergeometric techniques to a large class of Seiberg dualities.}
Coincidently, our integral~\eqref{indexWeak} satisfies the above
requirement and applying the theorem we can transform it into \be
I_{A_{2}}^{(2)}\left(1|t^{\frac{3}{2}}b{\mathbf
z},t^{\frac{3}{2}}a^{-1}{\mathbf
y}^{-1};t^{\frac{3}{2}}b^{-1}{\mathbf
z}^{-1},t^{\frac{3}{2}}a{\mathbf y}\right)=
I_{A_{2}}^{(2)}\left(1|t^{\frac{3}{2}}b^{-1}{\mathbf
z}^{-1},t^{\frac{3}{2}}a{\mathbf y}; t^{\frac{3}{2}}b{\mathbf
z},t^{\frac{3}{2}}a^{-1}{\mathbf y}^{-1}\right)\,. \ee Note that the
factor $\prod_{r,s=1}^{m+n+2}\Gamma(t_{r}u_{s})$ in~\eqref{Rains}
reduces to $1$ after pairwise cancelations using the
property~\eqref{inverse}. What we have effectively achieved through
this transformation is that we have exchanged the $U(1)$ quantum
numbers of the matter charged under the $SU(3)^{2}$ flavor. This in
particular implies that both the $SU(3)$ flavor groups are on the
same footing and are not associated with separate $U(1)$'s.

\subsection{Strongly-coupled frame and the index of $E_6$ SCFT}\label{strongSec}

\begin{figure}
\begin{centering}
\includegraphics[scale=0.7]{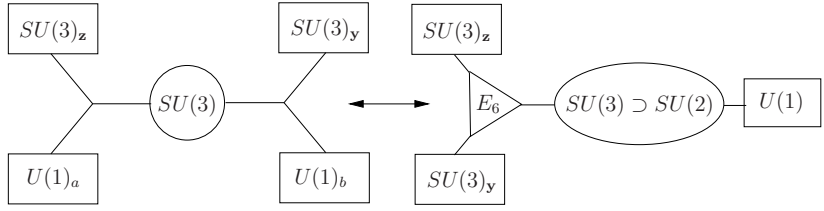}
\par\end{centering}

\caption{\label{strongFig} Argyres-Seiberg duality for $SU(3)$ SYM with $N_f = 6$.}

\end{figure}

In the strongly-coupled S-duality frame, figure~\ref{strongFig}, we have a
fundamental hypermultiplet coupled to an $SU(2)$ gauge theory. This
gauge group is identified with an $SU(2)$ subgroup of the $E_{6}$ flavor symmetry of a strongly-coupled rank one SCFT.
We do not know the field content of the strongly-coupled rank $1$ $E_{6}$ SCFT.
 This implies that we
can not write down the ``single letter'' partition function for that theory
and, a-priori, can not directly compute its index. In what follows we will use the index
computed in the weakly-coupled frame~\eqref{indexWeak} and the above statements about  Argyres-Seiberg duality
to infer the index of the $E_{6}$ SCFT.

\

Let $C^{(E_6)}$ denote the index of  rank $1$ $E_6$ SCFT~\cite{Minahan:1996fg}.
The maximal subgroup of $E_6$ is $SU(3)^{3}$. Two
among these three $SU(3)$'s are identified  with the two $SU(3)$ factors in the flavor group
of the weakly-coupled theory, see figure~\ref{strongFig}. Let the additional $SU(3)$ be denoted
by  $\mathbf{w}$. The fundamental representation of $E_6$ is
decomposed under $SU(3)_\mathbf{w}\otimes SU(3)_\mathbf{y}\otimes
SU(3)_\mathbf{z}$ as,
\begin{equation}
  \mathbf{27}_{E_6}
  =(\mathbf{3},\mathbf{\bar3},\mathbf{1})\oplus(\mathbf{\bar3},\mathbf{1},\mathbf{3})
   \oplus(\mathbf{1},\mathbf{3},\mathbf{\bar3})\,.
\end{equation}
Thus, the character of the $E_6$ fundamental fields is,
\begin{equation}
    \chi_{\mathbf{27}}
              =\sum_{i,j=1}^3\left(\frac{w_i}{y_j}+\frac{z_i}{w_j}+\frac{y_i}{z_j}\right)\,,\qquad
\prod_{i=1}^3 y_{i}=\prod_{i=1}^3 z_{i}=\prod_{i=1}^3 w_{i}=1\,.
\end{equation} The index $C^{(E_6)}$ is thus a function of ${\mathbf w},\,{\mathbf y}$, and ${\mathbf z}$.
 The S-duality picture suggests
that we should decompose $SU(3)_{\mathbf{w}}$ as $SU(2)_{e}\otimes U(1)_{r}$.
This amounts to the change of variables
$\{w_{1},w_{2},w_{2}\}\rightarrow\{er,\frac{r}{e},\frac{1}{r^{2}}\}$, for which
the character of the fundamental of $E_6$ becomes
\be
  \label{eq:character:E6:27}
    \chi_{\mathbf{27}}=(er+\frac{r}{e}+\frac{1}{r^2})(\frac{1}{y_1}+\frac{1}{y_2}+\frac{1}{y_3})
               +(\frac{1}{er}+\frac{e}{r}+r^2)(z_1+z_2+z_3)
               +\sum_{i,j=1}^3\frac{y_i}{z_j}\,.\nonumber\\
\ee
Thus, the index of the $E_6$ SCFT can be denoted as $C^{(E_6)}\left((e,\,r),\mathbf{y},\mathbf{z}\right)$.
In the above notations the index of the additional hypermultiplet of the theory is
\be
C_{s,\,e}=\Gamma\left(\frac{t^{2}}{\sqrt{v}}\,e{}^{\pm1}\,s^{\pm1}\right)\,.
\ee

Thus, one can write the superconformal index of the theory in the strongly-coupled frame as
\be\label{strong}
\hat{\mathcal I}\left(s,r;\mathbf{y},\mathbf{z}\right)
&=&{C_{s}}^e\,C^{(E_6)}_{(e,r),{\mathbf y},{\mathbf z}}=\\
 & = & \kappa\,\Gamma(t^{2}v)\;
\oint_{\mathbb T}\frac{de}{2\pi i\,e}\frac{\Gamma(t^{2}ve^{\pm2})}{\Gamma(e^{\pm2})}\Gamma(\frac{t^{2}}{\sqrt{v}}e^{\pm1}\,s^{\pm1})\;C^{(E_6)}\left((e,\,r),\mathbf{y},\mathbf{z}\right)\,.\nonumber
\ee
By Argyres-Seiberg duality we have to equate
\be\label{ASdual}
\hat{\mathcal I}\left(s,r;\mathbf{y},\mathbf{z}\right)=\mathcal{I}_{a,{\mathbf z};b,{\mathbf y}}\, ,
\ee where $\mathcal{I}_{a,{\mathbf z};b,{\mathbf y}}$ is given in~\eqref{indexWeak}, and we appropriately identify the $U(1)$ charges,
\be\label{U1s}
s=(a/b)^{3/2},\qquad r=\left(a\,b\right)^{-1/2}\,.
\ee

It so happens that the integral of equation~\eqref{strong} has
special properties which allow us to invert it (see appendix~\ref{inversionSec} and~\cite{spirinv}
for the details).
 One can write the following
\be\label{inverted} \kappa\oint_{C_w}\,\frac{d s}{2\pi
i\,s}\,\frac{\Gamma(\frac{\sqrt{v}}{t^2}w^{\pm1}\,s^{\pm1})}{\Gamma(\frac{v}{t^4},\;s^{\pm2})}\,
\hat{\mathcal I}\left(s,r;\mathbf{y},\mathbf{z}\right)
&=&\Gamma(t^{2}v\,w^{\pm2})\;C^{(E_6)}\left((w,\,r),\mathbf{y},\mathbf{z}\right)\,
, \ee where the contour $C_w$ is a deformation of the unit circle
such that it encloses $s=\frac{\sqrt{v}}{t^2}\,w^{\pm1}$ and
excludes $s=\frac{t^2}{\sqrt{v}}\,w^{\pm1}$ (for precise definition
and details see appendix~\ref{inversionSec} and~\cite{spirinv}).
The above expression for the index $C^{(E_6)}$ does satisfy~\eqref{strong}, but \textit{a-priori}  does not uniquely follow from it. 
However, as we will explicitly see below,~\eqref{inverted} is consistent with what is expected from $E_6$ SCFT.
We will comment on this issue in the end of this section.
 We can thus use the Argyres-Seiberg duality~\eqref{ASdual} to write  a closed form
expression for the $E_6$ index
\begin{equation}\label{strongIndex}
\addtolength{\fboxsep}{5pt}
\boxed{
\begin{gathered}
 C^{(E_6)}\left((w,\,r),\mathbf{y},\mathbf{z}\right)=\frac{2\,\kappa^3\Gamma(t^{2}v)^{2}}{3\,\Gamma(t^{2}v\,w^{\pm2})}\;
\oint_{C_w}\,\frac{d s}{2\pi i\, s}\,\frac{\Gamma(\frac{\sqrt{v}}{t^2}w^{\pm1}\,s^{\pm1})}{\Gamma(\frac{v}{t^4},\;s^{\pm2})}\,\times
\\
\times\;\oint_{\mathbb{T}^{2}}\prod_{i=1}^{2}\frac{dx_{i}}{2\pi i \,x_{i}}
\frac{\displaystyle\prod_{i=1}^{3}\prod_{j=1}^{3}\Gamma\left(\frac{t^{2}}{\sqrt{v}}\left(\frac{s^{\frac{1}{3}}\,z_{i}}{x_{j}\,r}\right)^{\pm1}\right)
\Gamma\left(\frac{t^{2}}{\sqrt{v}}\left(\frac{s^{-\frac{1}{3}}\,y_{i}\,x_{j}}{r}\right)^{\pm1}\right)
\prod_{i\neq j}\Gamma\left(t^{2}v\frac{x_{i}}{x_{j}}\right)}{\displaystyle\prod_{i\neq j}\Gamma\left(\frac{x_{i}}{x_{j}}\right)}\,.
\end{gathered}
}
\end{equation}
One can rewrite the above expression without using the special integration contour.
The integration contour $C_w$ can be split into five pieces: a contour around the unit circle $\mathbb{T}$, two contours encircling 
the simple poles of $\Gamma(\frac{\sqrt{v}}{t^2}w^{\pm1}\,s^{\pm1})$ at $s=\frac{\sqrt{v}}{t^2}w^{\pm1}$, and 
two contours encircling in the opposite direction
the simple poles of $\Gamma(\frac{\sqrt{v}}{t^2}w^{\pm1}\,s^{\pm1})$ at $\frac{t^2}{\sqrt{v}}w^{\pm1}$.
 Using the fact that elliptic Gamma function
satisfies~\eqref{limGamm} we have
\be
&&C^{(E_6)}\left((w,\,r),\mathbf{y},\mathbf{z}\right)=
\frac{\kappa}{\Gamma(t^2vw^{\pm2})}\oint_\mathbb{T}\,\frac{d
s}{s}\,\frac{\Gamma(\frac{\sqrt{v}}{t^2}w^{\pm1}\,s^{\pm1})}{\Gamma(\frac{v}{t^4},\;s^{\pm2})}\,
\hat{\mathcal I}\left(s,r;\mathbf{y},\mathbf{z}\right)\\
&&\qquad+\frac{1}{2}\frac{\Gamma(w^{-2})}{\Gamma(t^2vw^{-2})}\[\hat{\mathcal
I}\(s=\frac{\sqrt{v}w}{t^2},r;\mathbf{y},\mathbf{z}\)+\hat{\mathcal
I}\(s=\frac{t^2}{\sqrt{v}w},r;\mathbf{y},\mathbf{z}\)\]\nonumber\\
&&\qquad+\frac{1}{2}\frac{\Gamma(w^{2})}{\Gamma(t^2vw^{2})}\[\hat{\mathcal
I}\(s=\frac{\sqrt{v}}{t^2w},r;\mathbf{y},\mathbf{z}\)+\hat{\mathcal
I}\(s=\frac{t^2w}{\sqrt{v}},r;\mathbf{y},\mathbf{z}\)\]\,.\nonumber\ee

The index~\eqref{strongIndex} encodes some information about the
matter content of the $E_6$ theory.
To extract this information it is useful to expand the index~\eqref{strongIndex} in the chemical potentials.
We define an expansion in $t$ as
\be
C^{(E_6)}\equiv \sum_{k=0}^\infty a_k\, t^k \,.
\ee
The first several orders in this expansion have the following form
    \begin{equation}\label{eq:E6expansion}
      \begin{split}
        a_0=&1\\
        a_1t=&a_2t^2=a_3t^3=0\\
        a_4t^4=&\frac{t^4}{v}\chi^{E_6}_{\mathbf{78}}\\
        a_5t^5=&0\\
        a_6t^6=&-t^6\chi^{E_6}_{\mathbf{78}}-t^6+t^6v^3\\
        a_7t^7=&\frac{t^7}{v}\left(y+\frac{1}{y}\right)\chi^{E_6}_{\mathbf{78}}+\frac{t^7}{v}\left(y+\frac{1}{y}\right)
               -t^7v^2\left(y+\frac{1}{y}\right)\\
        a_8t^8=&\frac{t^8}{v^2}\left(\chi^{E_6}_{\sym{2}{\rep{78}}}-\chi^{E_6}_\rep{650}-1\right)
                +t^8v+t^8v\\
        a_9t^9=&-t^9\left(y+\frac{1}{y}\right)\chi^{E_6}_\rep{78}-2t^9\left(y+\frac{1}{y}\right)
                +t^9v^3\left(y+\frac{1}{y}\right)\\
a_{10}t^{10}=&-\frac{t^{10}}{v}(\chi^{E_6}_{\rep{78}}\,\chi^{E_6}_{\rep{78}}-\chi^{E_6}_\rep{650}-1)
  +\frac{t^{10}}{v}\(y^2+1+\frac{1}{y^2}\)\chi^{E_6}_\rep{78}+\\
  &+\frac{t^{10}}{v}\(y+\frac{1}{y}\)^2-t^{10}v^2\(y+\frac{1}{y}\)^2\\
        a_{11}t^{11}=&\frac{t^{11}}{v^2}\(y+\frac{1}{y}\)(\chi^{E_6}_{\rep{78}}\,\chi^{E_6}_{\rep{78}}-\chi^{E_6}_\rep{650}-1)
                      +t^{11}v\(y+\frac{1}{y}\)+t^{11}v\(y+\frac{1}{y}\)\,.
      \end{split}
    \end{equation}
The adjoint representation of $E_6$ , $\mathbf{78}$, decomposes in
the following way in terms of its maximal $SU(3)^3$ subgroup \be
\mathbf{78}&=&(\mathbf{3},\mathbf{3},\mathbf{3})+(\mathbf{\bar{3}},\mathbf{\bar{3}},\mathbf{\bar{3}})+
(\mathbf{8},\mathbf{1},\mathbf{1})+(\mathbf{1},\mathbf{8},\mathbf{1})+(\mathbf{1},\mathbf{1},\mathbf{8})\,,
\ee and $\mathbf{650}$ of $E_6$ is composed as \be
\mathbf{650}=\mathbf{27}\times\mathbf{\overline{27}}-\mathbf{78}-\mathbf{1}\,.
\ee 
 The Higgs branch operators
$\mathbb{X}$ of $E_6$ theory are in the adjoint ($\rep{78}$)
representation of $E_6$ flavor algebra. The terms of the index
proportional to $\chi^{E_6}_{\rep{78}}$ are forming the following series,
\begin{equation}
  \[\frac{t^4}{v}-t^6
  +\frac{t^7}{v}\left(y+\frac{1}{y}\right)
  -t^9\left(y+\frac{1}{y}\right)+\cdots\]\chi^{E_6}_\rep{78}\,,
\end{equation}
which is the index of a multiplet with $\Delta=2$, $j=\bar{j}=0$ and $r=0$ and of its derivatives
 (see appendix C.2 of
\cite{Gadde:2009dj}). Taken as a ``letter'' this multiplet has the following  ``single letter'' partition function
\begin{equation}
  \frac{t^4/v-t^6}{(1-t^3y)(1-t^3/y)}\,,
\end{equation}
which matches the quantum numbers of the Higgs branch operators on the
weakly-coupled  side of the Argyres-Seiberg duality if we follow the identifications listed in
\cite{Gaiotto:2008nz}.

The $E_6$ singlet part of the index contains yet another series,
\begin{equation}
  t^6v^3-t^7v^2\(y+\frac{1}{y}\)+t^8v+t^9v^3\(y+\frac{1}{y}\)+\cdots\,.
\end{equation}
This series forms the index of a chiral multiplet with $\Delta=3$,
$j=\bar{j}=0$ and $r=3$ together with its derivatives (appendix C.1
of \cite{Gadde:2009dj})
\begin{equation}
  \frac{t^6v^3-t^7v^2\(y+\frac{1}{y}\)+t^8v}{(1-t^3y)(1-t^3/y)}\,.
\end{equation}
Since the Coulomb branch operator, $u$, of $E_6$ theory (which is
identified as $\tr\phi^3$ of the dual $SU(3)$ theory) has exactly
the same quantum numbers, this multiplet is identified as the Coulomb
branch operator.

The remaining singlet part of the index,
\begin{equation}
  -t^6+\frac{t^7}{v}\(y+\frac{1}{y}\)+t^8v-2t^9\(y+\frac{1}{y}\)+\cdots\,,
\end{equation}
 is just the index of the stress tensor multiplet and its
derivatives  (appendix C.3 of \cite{Gadde:2009dj})
\begin{equation}
  \frac{-t^6+\frac{t^7}{v}\(y+\frac{1}{y}\)+t^8v-t^9\(y+\frac{1}{y}\)}{(1-t^3y)(1-t^3/y)}\,.
\end{equation}

Besides the matter content, the index also provides possible
constraints among operators. For example, it was argued~\cite{Gaiotto:2008nz} that the Higgs branch operators
of the $E_6$ theory should obey
the Joseph relations,
\begin{equation}
  (\mathbb{X}\otimes\mathbb{X})|_{\mathcal{I}_2}=0\,,
\end{equation}
where the representation $\mathcal{I}_2$ is defined as
\begin{equation}
  \sym{2}{V(\rep{adj})}=V(2\rep{adj})\oplus\mathcal{I}_2\,.
\end{equation}
For $E_6$, $\rep{adj}=\rep{78}$, $2\rep{adj}=\rep{2430}$ and then
$\sym{2}{\rep{78}}=\rep{2430}\oplus\rep{650}\oplus\rep{1}$. Thus, in our case
\begin{equation}
  \mathcal{I}_2=\rep{650}\oplus\rep{1}\,.
\end{equation}
The Joseph relation in $E_6$ theory reads,
\begin{equation}
  \label{eq:josephrelation}
  (\mathbb{X}\otimes\mathbb{X})|_{\rep{650}\oplus\rep{1}}=0\,,
\end{equation}
which means that these operators should not appear in the index.
 The index of $\mathbb{X}$ is $t^4/v$, then the index of
$\mathbb{X}\otimes\mathbb{X}$ is $t^8/v^2$. \eqref{eq:E6expansion}
shows that our index is consistent with the Joseph relation.

Further constraints can also be derived from the higher order terms in~\eqref{eq:E6expansion}.
Let us consider the index at order $t^{10}$.
The meaning of each term is clear. The first term corresponds to
operators $\mathbb{X}\otimes(Q\mathbb{X})$ with the constraint
$Q(\mathbb{X}\otimes\mathbb{X})_{\rep{650}+1}=0$ which is a descendant
of Joseph relation above~\eqref{eq:josephrelation}. The last three
terms are derivative descendants of
$\frac{t^4}{v}\chi^{E_6}_\rep{78}$,
$\frac{t^{7}}{v}\left(y+\frac{1}{y}\right)$ and
$-t^{7}v^2\left(y+\frac{1}{y}\right)$
respectively. However, terms of the form
\begin{equation}
  t^{10}v^2\chi^{E_6}_\rep{78}\,,
\end{equation}
which would be corresponding to the $Higgs\otimes Coulomb$ operators are absent.
This fact implies the constraint
\begin{equation}
  \mathbb{X}\otimes u=0\,.
\end{equation}
This is consistent with the fact that the $E_6$ theory has rank 1.
The absence of $-\frac{t^{10}}{v}\chi^{E_6}_\rep{78}$ also implies the
constraint
\begin{equation}
  \mathbb{X}\otimes T=0\,,
\end{equation}
where $T$ is the stress tensor. The structure of the index at order $t^{11}$ is
 consistent with these two constraints.

Finally, let us comment on the uniqueness of our proposal.
 In principle, 
 the index~\eqref{strongIndex} produced by the construction of this section might  differ from
the true index of the $E_6$ SCFT: $C^{(E_6)}_{true}((e,\,r),\mathbf{y},\mathbf{z})
= C^{(E_6)}((e,\,r),\mathbf{y},\mathbf{z})+\delta C((e,\,r),\mathbf{y},\mathbf{z})$, with $\delta C$ satisfying
\be
\oint_{\mathbb{T}}\frac{de}{2\pi i\, e}\frac{\Gamma(\frac{t^2}{\sqrt{v}}\,e^{\pm1}s^{\pm1})
\Gamma(t^2v\,e^{\pm2})}{\Gamma(e^{\pm2})}\,\delta C((e,\,r),\mathbf{y},\mathbf{z})=0\,. 
\ee
At this stage we are not able to rigorously rule out such a possibility. However, the $E_6$ covariance of our proposal,
its consistency with physical expectations about protected operators and the further S-duality checks
performed in the following section, make us confident that we have identified the
 correct index  of the $E_6$ SCFT.

Note that the expression for the index~\eqref{strongIndex} is
not explicitly given in terms of $E_6$ characters. However, as one
learns from the perturbative expansion~\eqref{eq:E6expansion}, the characters of
$SU(3)_\mathbf{y}\otimes SU(3)_\mathbf{z}\otimes SU(2)_w\otimes U(1)_r$
always  combine into $E_6$ characters. Essentially, since the weakly-coupled frame has really $SU(6)\otimes U(1)$ flavor symmetry we can
write an expression for the $E_6$ index which has a manifest
$SU(6)\otimes SU(2)$ symmetry,\footnote{The fact that this symmetry
can be manifestly seen in the expression for the index is very
reminiscent of the construction of the $E_6$ symmetry using
multi-pronged strings in~\cite{Gaberdiel:1997ud}. It is very
interesting to understand whether these facts are related.} but not
the full $E_6$. The fact that just by assuming Argyres-Seiberg duality we obtain an index
for a theory with an $E_6$ flavor symmetry and with a consistent spectrum of operators
  is a non-trivial check of Argyres-Seiberg duality.

\section{S-duality checks of the $E_6$ index}\label{sdualSec}

In the previous section we have discussed the superconformal index
of the $N_f=6$ $SU(3)$ theory and of its strongly-coupled dual. One
can obtain this theory by compactifying a $(2,0)$ $6d$ theory on a
sphere with four punctures, two $U(1)$ punctures and two $SU(3)$
punctures. The different S-duality frames are then given by the
different degeneration limits of this Riemann surface. The weakly-coupled frames are obtained by bringing together one of the $U(1)$
punctures and one of the $SU(3)$ punctures, and the strongly-coupled
frame is obtained by colliding the two $SU(3)$ ($U(1)$) punctures.
The coupling constant of the theory is related to the cross ratio of
the four punctured sphere.

In~\cite{Gaiotto:2009we} Gaiotto suggested to generalize this
picture by considering general Riemann surfaces with an arbitrary
numbers of punctures of different types (two types in case of the
$SU(3)$ theories). The claim is that all theories with the same
number and type of punctures and same topology of the Riemann
surface are related by S-dualities. The immediate consequence of
this claim for the superconformal index is that all such theories
have to have the same index as it is independent of the values of
the coupling, i.e. the moduli of the Riemann surface. This implies
that the superconformal index is a topological invariant of the
punctured Riemann surface. It was claimed in~\cite{Gadde:2009kb}
that the superconformal index can be actually interpreted as a
correlator in a two  dimensional topological quantum field theory.
The structure constants of this TQFT are given by the index of the
three punctured sphere and the contraction of indices (i.e. metric)
is gauging of the flavor symmetries. The associativity of the
algebra generated by the structure constants is equivalent to the
invariance of the index of four punctured spheres under
pair-of-pants decomposition into two three punctured spheres. The
structure constants and  the metric were constructed and the
associativity was explicitly verified for the $SU(2)$ case.

In this section we will make the same analysis for the $SU(3)$ case.
We have two types of punctures, associated to $U(1)$ and $SU(3)$
flavor symmetries. There are thus different three point functions
one can construct. The index of the theory on a sphere with three
$SU(3)$ punctures, i.e. the index of the $E_6$ theory, is a
structure constant which we will denote by $C^{(333)}_{{\mathbf
x},{\mathbf y},{\mathbf z}}$ and it is just given by
\eqref{strongIndex},
\begin{equation}\label{rank1}
  C^{(333)}_{\mathbf{x,y,z}}=C^{(E_6)}\(\left(\sqrt{\frac{x_1}{x_2}},\sqrt{x_1x_2}\right),\mathbf{y},\mathbf{z}\)\,.
\end{equation}
This vertex corresponds to the $E_6$ theory which has rank one, and
thus we will refer to it as a rank ${\bf 1}$ vertex. We will denote
by $C^{(133)}_{{\mathbf x},{\mathbf y},a}$ the index of the sphere
with two $SU(3)$ punctures and one $U(1)$ puncture. This is a free
theory consisting of a hypermultiplet in fundamental of two $SU(3)$
flavor groups and its value is given by~\eqref{vert331},
\begin{equation}\
  \label{rank0}
  C^{(133)}_{a,\mathbf{x},\mathbf{y}}=\prod_{i,j=1}^3\Gamma\left(\frac{t^2}{\sqrt{v}}\(ax_iy_j\)^\pm\right)\,.
\end{equation}
 This vertex
 corresponds to a free, rank ${\bf 0}$, theory and we will refer to it as rank zero structure constant.
Later on we will define yet another three point function, formally associated to a sphere with two $U(1)$ punctures
and one $SU(3)$ puncture. This vertex will have effective rank ${\bf -1}$.
The metric of the model, $\eta^{\mathbf{x},\mathbf{y}}$, is defined as
\begin{equation}
  \label{eq:def:metric}
  \eta^{\mathbf{x},\mathbf{y}}=\frac{2}{3}\,\kappa^2\,\Gamma^2(t^2v)
                     \prod_{1\leqslant
                     i<j\leqslant3}\frac{\Gamma\(t^2v\(\frac{x_i}{x_j}\)^\pm\)}{\Gamma\(\(\frac{x_i}{x_j}\)^\pm\)}
                     \hat\Delta(\mathbf{x}^{-1},\mathbf{y})\,,
\end{equation}
where $\hat\Delta(\mathbf{x}^{-1},\mathbf{y})$ is a $\delta$-function kernel defined by
\be\label{kernelSu3}
\oint_{\mathbb{T}^2}\prod_{i=1}^2\frac{dx_i}{2\pi i\,x_i}\,\hat\Delta(\mathbf{x},\mathbf{w})\, f(\mathbf{x})=f(\mathbf{w})\,,\qquad
\mathbf{w}\in \mathbb{T}^2\,.
\ee
The indices are contracted as follows
\begin{equation}
  A^{\ldots\mathbf{u}\ldots}B_{\ldots\mathbf{u}\ldots}
  \equiv\left.\oint_{\mathbb{T}^2}\prod_{i=1}^2\frac{du_i}{2\pi
  iu_i}\,A^{\ldots\mathbf{u}\ldots}B_{\ldots\mathbf{u}\ldots}\right|_{\prod_{i=1}^3u_i=1}\,.
\end{equation}
Following these definitions the superconformal indices of all the
$SU(3)$ generalized quivers are obtained by contracting the
structure constants in different ways.

For the S-duality to hold, and subsequently for the structure constants to have a TQFT interpretation, the
algebra generated by these objects has to be associative. We proceed to verify this fact.
\begin{figure}[htbp]
\begin{center}
$\begin{array}{c@{\hspace{0.75in}}c@{\hspace{0.75in}}c}
\epsfig{file=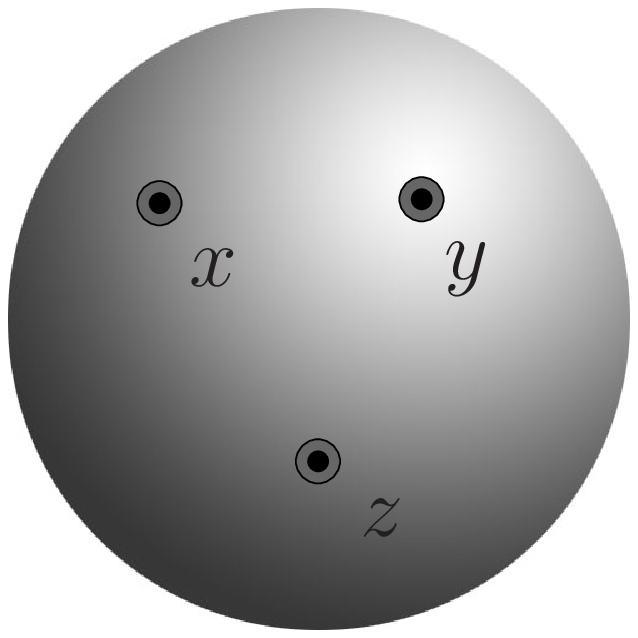,scale=0.4} & \epsfig{file=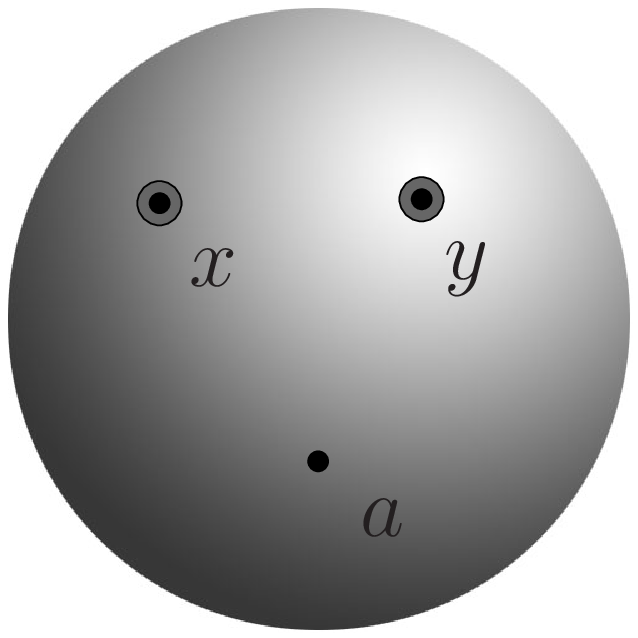,scale=0.4}& \epsfig{file=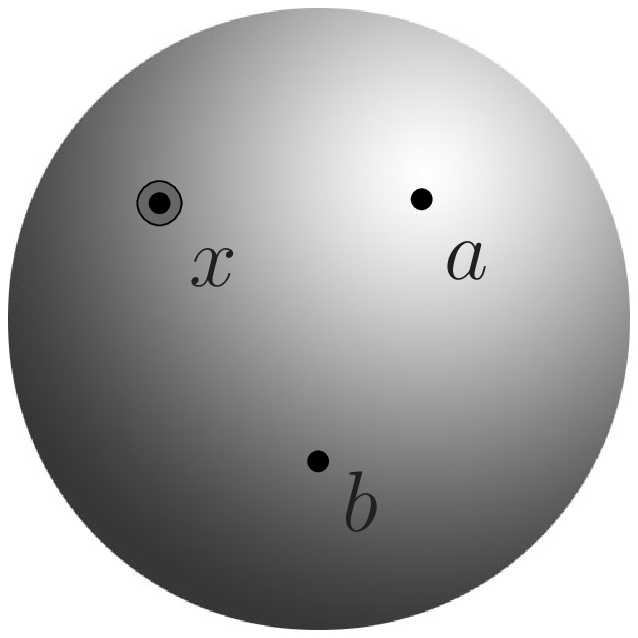,scale=0.4}\\\
 C^{(333)}_{\mathbf{x,y,z}}&  C^{(133)}_{a,\mathbf{x},\mathbf{y}} &  C^{(113)}_{a,b,\mathbf{x}}\\ [0.2cm]
\end{array}$
\end{center}
 \begin{center}
\caption{The three structure constants of the TQFT. The dots represent
 $U(1)$ punctures and the circled dots  $SU(3)$ punctures.
\label{vertices} }
\end{center}
\end{figure}

\subsection*{$(333)-(333)$ associativity}
Let us consider the generalized quiver with genus zero and four $SU(3)$ punctures.
The index should be invariant under the permutation of the four $SU(3)$ characters,
\begin{eqnarray}
  \label{eq:duality(0,4)0}
  \mathcal{I}_{3333}(\mathbf{x},\mathbf{y};\mathbf{w},\mathbf{z})
  &=&C^{(333)}_{\mathbf{x,y,u}}\eta^{\mathbf{u,v}}C^{(333)}_{\mathbf{v,z,w}}
  =C^{(333)}_{\mathbf{x,z,u}}\eta^{\mathbf{u,v}}C^{(333)}_{\mathbf{v,y,w}}\,.
\end{eqnarray}
At order $O(t^4)$ we find ,
\begin{equation}
  \mathcal{I}_{3333}\sim
  t^4\[\frac{1}{v}(\chi_\rep{8}(\mathbf{x})+\chi_\rep{8}(\mathbf{y})+\chi_\rep{8}(\mathbf{z})+\chi_\rep{8}(\mathbf{w}))
       +v^2\]\,,
\end{equation}
and at order $O(t^6)$,
\begin{equation}
  \mathcal{I}_{3333}\sim
  t^6\[-(\chi_\rep{8}(\mathbf{x})+\chi_\rep{8}(\mathbf{y})+\chi_\rep{8}(\mathbf{z})+\chi_\rep{8}(\mathbf{w}))
       +3v^3\]\,.
\end{equation}
These axpressions are symmetric under the exchange
$\mathbf{x}\leftrightarrow\mathbf{y}\leftrightarrow\mathbf{z}\leftrightarrow\mathbf{w}$.
The associativity can be checked to hold to higher orders as well.

\subsection*{$(333)-(331)$ associativity}
Let us consider the generalized quiver with genus zero, three $SU(3)$ punctures and one $U(1)$ puncture.
The index should be invariant under permutations of the three $SU(3)$ characters
\begin{eqnarray}
  \label{eq:duality(1,3)0}
  \mathcal{I}_{3331}(a,\mathbf{x};\mathbf{y},\mathbf{z})
  &=&C^{(133)}_{a,\mathbf{x,u}}\eta^{\mathbf{uv}}C^{(333)}_{\mathbf{v,y,z}}
  =C^{(133)}_{a,\mathbf{y,u}}\eta^{\mathbf{uv}}C^{(333)}_{\mathbf{v,x,z}}\,.
\end{eqnarray}
We also expand the integrand in $t$ around $t=0$. The first
non-trivial check is for the coefficient of $\mathcal{I}_{3331}$ at order
$O(t^4)$,
\begin{equation}
  \mathcal{I}_{3331}\sim t^4\[\frac{1}{v}\(\chi_\rep{8}(\mathbf{x})+\chi_\rep{8}(\mathbf{y})+\chi_\rep{8}(\mathbf{z})+1\)+v^2\]\,,
\end{equation}
which is indeed symmetric under
$\mathbf{x}\leftrightarrow\mathbf{y}\leftrightarrow\mathbf{z}$. At
order $O(t^6)$,
\begin{eqnarray}
  \mathcal{I}_{3331}&\sim&\frac{t^6}{v^{3/2}}\(a^{-3}+a^{-1}\chi_{\crep{3}}(\mathbf{x})\chi_{\crep{3}}(\mathbf{y})\chi_{\crep{3}}(\mathbf{z})
                   +a\chi_\rep{3}(\mathbf{x})\chi_\rep{3}(\mathbf{y})\chi_\rep{3}(\mathbf{z})+a^3\)\\
             &&-t^6\(\chi_\rep{8}(\mathbf{x})+\chi_\rep{8}(\mathbf{y})+\chi_\rep{8}(\mathbf{z})+1\)+2t^6v^3\,,\nonumber
\end{eqnarray}
which is also symmetric under
$\mathbf{x}\leftrightarrow\mathbf{y}\leftrightarrow\mathbf{z}$. Again, we
can perform systematic checks  to arbitrary high order in $t$.

\subsection*{The $(311)$ three point function and $(311)-(331)$ associativity}
The index of the $N_f=6$ $SU(3)$ theory in the strongly-coupled frame is given in terms of
an integral over an $SU(2)$ character. Thus, we can not write it using the structure constants
and the metric we defined in the beginning of this section. The strongly-coupled frame is obtained when two
$U(1)$ punctures collide and thus in what follows we will formally define a structure constant with two $U(1)$ characters
and an $SU(3)$ character such that when contracted with the $E_6$ structure constant using the metric above
it will produce the index of the strongly-coupled frame.

Let us rewrite the index in the strongly-coupled frame,
\be
\hat{\mathcal I}\left(s,r;\mathbf{y},\mathbf{z}\right)
 & = & \kappa\;\Gamma(t^2 v)\,
\oint_{\mathbb{T}}\frac{de}{2\pi i\,e}\frac{\Gamma(\frac{t^{2}}{\sqrt{v}}e^{\pm}\,s^\pm)}{\Gamma(e^{\pm2})}\;\Gamma(t^{2}v\,e^{\pm2})\;C\left((e,\, r),\mathbf{y},\mathbf{z}\right)\,,
\ee as rank one ($E_6$) $(333)$  and rank $-1$ $(113)$ vertices contracted
\be
\hat{\mathcal I}\left(a,b;\mathbf{y},\mathbf{z}\right)&=&C^{(113)}_{a,b,\mathbf{x}}\,\eta^{\mathbf{x},\mathbf{x'}}\,C^{(333)}_{\mathbf{x'},
\mathbf{y},\mathbf{z}}=\\
 & = & \frac{2}{3}\,\kappa^2\,\Gamma(t^2v)^2\;
\oint_{\mathbb{T}^2}\prod_{i=1}^2\frac{dx_i}{2\pi i\,x_i}\prod_{i\neq j}\frac{\Gamma(t^{2}v\,x_i/x_j)}{\Gamma(x_i/x_j)}\;
C^{(113)}\left(a,b,\mathbf{x}^{-1}\right)\;C^{(333)}\left(\mathbf{x},\mathbf{y},\mathbf{z}\right)\, .\nonumber
\ee For this we define
\be\label{Vert113}
C^{(113)}\left(a,b,\mathbf{x}^{-1}\right)&=&\frac{3}{2\kappa\Gamma(t^2v)}\,
\oint_{\mathbb{T}}\frac{de}{2\pi i\,e}\frac{\Gamma(\frac{t^{2}}{\sqrt{v}}\,e^{\pm1}\,s^{\pm1})\,\Gamma(t^{2}v\,e^{\pm2})}{\Gamma(e^{\pm2})}
\prod_{i\neq j}\frac{\Gamma(x_i/x_j)}{\Gamma(t^{2}v\,x_i/x_j)}\, \hat\Delta(\mathbf{x},\mathbf{w})\,.\nonumber\\
\ee
Here, $\mathbf{w}=(e,\, r)$ with $e$ an $SU(2)$ character and $r$ a $U(1)$ character. The $U(1)$ charges are related as in~\eqref{U1s},
 $s=(a/b)^{3/2}$ and $r=(a\,b)^{-1/2}$.
 $\hat \Delta(\mathbf{x},\mathbf{w})$ is a $\delta$-function kernel defined in~\eqref{kernelSu3}.
The $(113)$ vertex has effective rank ${\bf -1}$. Using the above definition the TQFT algebra is well
defined with all the contractions being $SU(3)$ integrals.

The associativity of $(311)$ vertex contracted with a $(333)$ vertex is achieved by construction: remember that we obtained the
index of $E_6$ SCFT by requiring this property.
Let us check the associativity of $(331)$ contracted with $(113)$
\be \label{su2}
&&{\mathcal I}(a,b;\,c,{\mathbf y})=C^{(113)}_{a,b,\mathbf{x}}\,\eta^{\mathbf{x},\mathbf{x'}}\,C^{(331)}_{\mathbf{x'},
\mathbf{y},c}=\\
&&\qquad \frac{2}{3}\,\kappa^2\,\Gamma(t^2v)^2\;
\oint\prod_{i=1}^2\frac{dx_i}{2\pi i\,x_i}\prod_{i\neq j}\frac{\Gamma(t^{2}v\,x_i/x_j)}{\Gamma(x_i/x_j)}\;
C^{(113)}\left(a,b,\mathbf{x}^{-1}\right)\;\prod_{i,j}\Gamma\left(\frac{t^2}{\sqrt{v}}\left(c\,x_i\,y_j\right)^{\pm1}\right)\, .\nonumber\\
&&\qquad\qquad\qquad\qquad=\prod_{i=1}^3\Gamma\left(\frac{t^2}{\sqrt{v}}\left(\frac{c\,y_i}{r^2}\right)^{\pm1}\right)\times\nonumber\\
&&\qquad\qquad\kappa\Gamma(t^2v)\oint\frac{de}{2\pi
i\,e}\frac{\Gamma(t^{2}v\,e^{\pm2})}{\Gamma(e^{\pm2})}
\Gamma\left(\frac{t^{2}}{\sqrt{v}}\,s^{\pm1}\,e^{\pm1}\right)\Gamma\left(\frac{t^2}{\sqrt{v}}\left(c\,r\,y_i\right)^{\pm1}e^{\pm1}\right)\,.
\nonumber \ee This is exactly the index of $SU(2)$ $N_f=4$ (the
fourth line in \eqref{su2}) with a decoupled hypermultiplet in the
fundamental of an $SU(3)$ flavor (the third line in \eqref{su2}).
Remembering~\eqref{U1s} and the results of~\cite{Focco,Gadde:2009kb}
it is easy to show that there is a permutation symmetry between the
three $U(1)$ punctures $a$, $b$ and $c$, \be
a\quad\leftrightarrow\quad b\quad\leftrightarrow\quad c\,. \ee

Using the definition~\eqref{Vert113} the index of a sphere with four
$U(1)$ punctures is singular. However, we do not have a physical
interpretation of this surface and it does not appear in any
decoupling limit of a physical theory. Thus, making sense of this
surface is not essential.

We have shown that the structure constants define an associative
algebra and thus define a TQFT. In particular the superconformal
index of theories with equal genus and equal number/type of
punctures is the same in agreement with S-duality.
\begin{figure}[htbp]
\begin{center}
$\begin{array}{c@{\hspace{0.35in}}c@{\hspace{0.35in}}c@{\hspace{0.35in}}c}
\epsfig{file=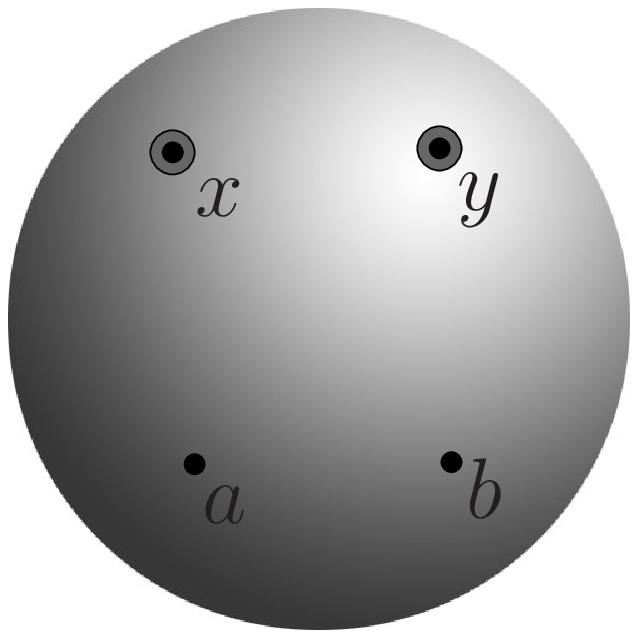,scale=0.4} & \epsfig{file=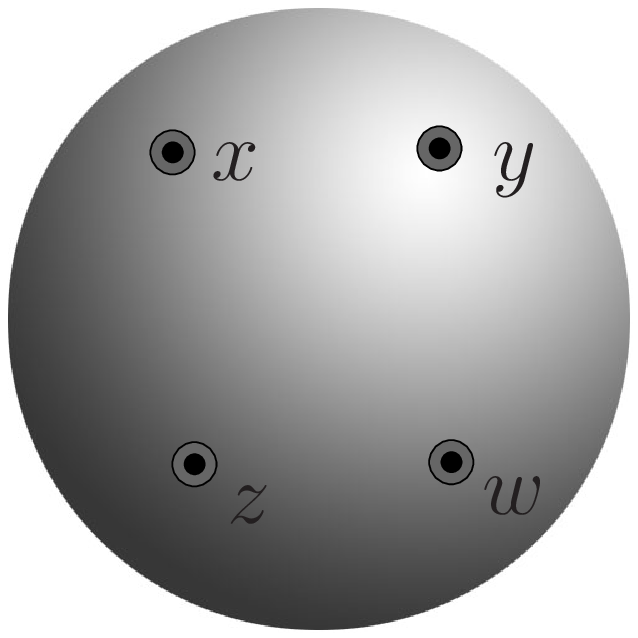,scale=0.4}& \epsfig{file=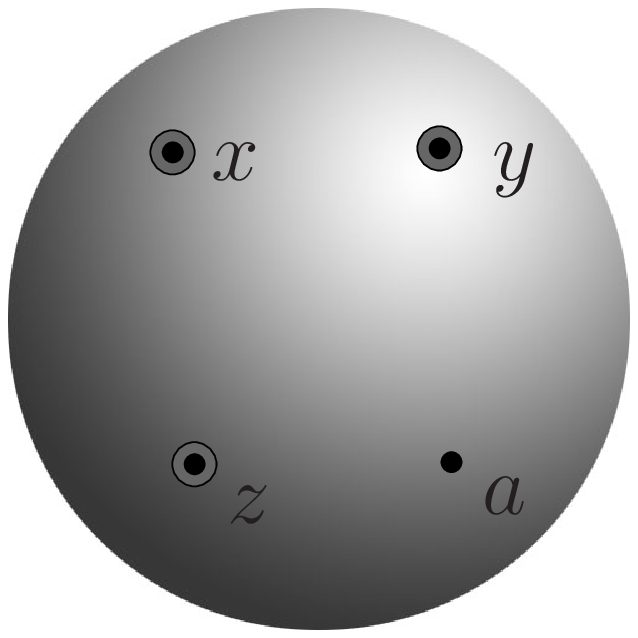,scale=0.4}& \epsfig{file=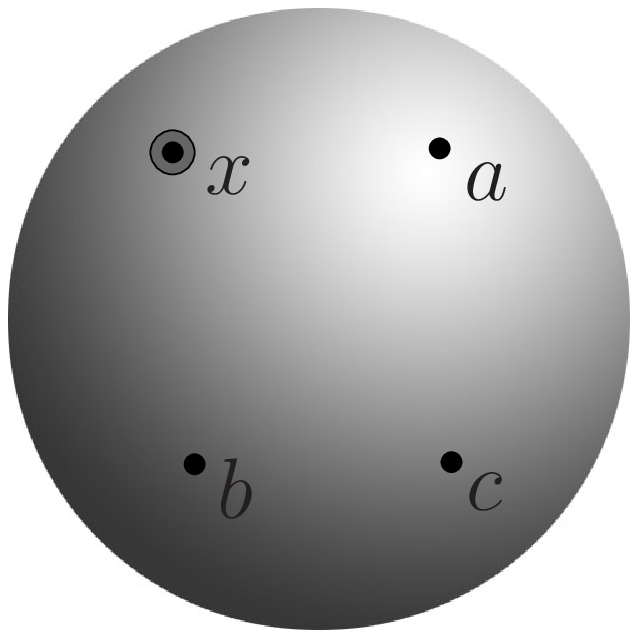,scale=0.4}\\\
 (a)&(b)&(c)&(d)\\ [0.2cm]
\end{array}$
\end{center}
 \begin{center}
\caption{The relevant four-punctured spheres for $A_2$ theories. The three different degeneration
limits of a four-punctured sphere correspond to different S-duality frames. For example, in
$(a)$ two of the degeneration limits (when a $U(1)$ puncture
 collides with an $SU(3)$ puncture)
 correspond to  the  weakly-coupled $N_f=6$ $SU(3)$ theory, the third limit (when two like punctures collide)
 corresponds to the Argyres-Seiberg theory.
%$(b)$ There is only  different degeneration limits correspond to permutations of the $SU(3)$ punctures.  $(c)$
%The crossing symmetry interchanges the three $SU(3)$ characters.
In $(d)$ the degeneration limits correspond to the different duality frames of
  $SU(2)$ SYM with $N_f=4$ theory plus a decoupled hypermultiplet. } 
  \label{channels}
\end{center}
\end{figure}

\section{Discussion}\label{discSec}
In this paper we have
obtained an explicit expression for the superconformal index of the
strongly-coupled SCFT with an $E_6$ flavor
symmetry~\cite{Minahan:1996fg}. The strategy is to use the
Argyres-Seiberg duality, which relates a weakly-coupled theory, index
of which can be easily obtained through the Lagrangian description
of the theory, and $E_6$ SCFT with part of the global symmetry
gauged. The index of the two theories should be the same. Thus, one
obtains the index of the $E_6$ theory by ``\textit{inverting}'' the
gauging, see~\eqref{strongIndex}. Upon gauging a flavor symmetry one
looses information about the theory by projecting on gauge invariant
states. However, what allows us to ``\textit{invert}'' the gauging
in our case is the fact that additional matter is coupled to the $SU(2)$ gauge group along with the $E_6$ SCFT, and
thus effectively preserves enough information to reconstruct the
complete index of $E_6$ SCFT. We do not have a physical
interpretation of the expression for the index~\eqref{strongIndex}
and it would be very interesting to find such an
interpretation.

\begin{figure}[htbp]
\begin{center}
\epsfig{file=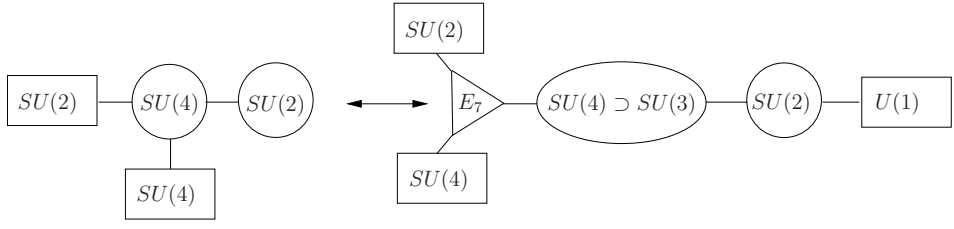,scale=0.7}
\end{center}
 \begin{center}
\caption{An Argyres-Seiberg duality relating a Lagrangian theory (left quiver) with a theory containing a strongly-coupled $E_7$ piece (right quiver).
} \label{E7}
\end{center}
\end{figure}
 In principle
one can try to use the same techniques to obtain the superconformal
index for other strongly-coupled SCFTs of~\cite{Gaiotto:2009we}.
However, the generalization is not completely straightforward. Let
us discuss the case of the $E_7$
theory~\cite{Minahan:1996cj,argyres-2007-0712,Benini:2009gi} as an
example. To obtain the $E_7$ SCFT we can apply  Argyres-Seiberg
duality to a Lagrangian theory with $SU(4)\otimes SU(2)$ gauge
group, with a single hypermultiplet in the bi-fundamental representation
and six hypermultiplets in the fundamental
representation of $SU(4)$. The Argyres-Seiberg dual of this theory
involves an $E_7$ strongly-coupled piece, with an $SU(3)$ subgroup of
$E_7$ gauged. The theory has a second gauge group factor $SU(2)$ and
two hypermultiplets: one in the fundamental of $SU(2)$ and the in
bi-fundamental of the two gauge groups.  See figure \ref{E7}. The
index of the weakly-coupled theory can be easily written down,
\be\label{E7weak} {\mathcal I}_{weak}&=&\kappa\,\Gamma(t^{2}v)\;
\oint_{\mathbb{T}}\frac{de}{2\pi i\,e}\frac{\Gamma(t^{2}ve^{\pm2})}{\Gamma(e^{\pm2})}\;\times\\
&&\frac{1}{3}\kappa^3\,\Gamma(t^{2}v)^3\oint_{\mathbb{T}^3}\prod_{i=1}^3\frac{du_i}{2\pi i\,u_i}\prod_{i\neq j}\frac{\Gamma(t^{2}v \frac{u_i}{u_j})}{\Gamma(\frac{u_i}{u_j})}\Gamma(\frac{t^{2}}{\sqrt{v}}(e^{\pm1}\,u_i\, a)^{\pm1})\,\times\nonumber\\
&&\prod_{i=1}^4\prod_{j=1}^4\Gamma(\frac{t^{2}}{\sqrt{v}}(y_j\,u_i\, b)^{\pm1})\,
\prod_{i=1}^4\prod_{j=1}^2\Gamma(\frac{t^{2}}{\sqrt{v}}(z_j\,u_i\, c)^{\pm1})\,.\nonumber
\ee
The index of the dual theory is given by
\be\label{E7strong}
{\mathcal I}_{strong}&=&\kappa\,\Gamma(t^{2}v)\;
\oint_{\mathbb{T}}\frac{de}{2\pi i\,e}\frac{\Gamma(t^{2}ve^{\pm2})}{\Gamma(e^{\pm2})}\Gamma(\frac{t^{2}}{\sqrt{v}}e^{\pm1}\,s^{\pm1})\;\times\\
&&\frac{2}{3}\kappa^2\,\Gamma(t^{2}v)^2\oint_{\mathbb{T}^2}\prod_{i=1}^2\frac{du_i}{2\pi i\,u_i}\prod_{i\neq j}\frac{\Gamma(t^{2}v \frac{u_i}{u_j})}{\Gamma(\frac{u_i}{u_j})}\prod_{i=1}^3\Gamma(\frac{t^{2}}{\sqrt{v}}(e^{\pm1}\,u_i\, m)^{\pm1})\;\times\nonumber\\
&&C^{(E_7)}\left((u_i,\,r)_{SU(4)},\mathbf{y}_{SU(4)},\mathbf{z}_{SU(2)}\right)\,.\nonumber
\ee One can invert the $SU(2)$ integral by the same techniques we
used for the $E_6$ index, but there is no simple inversion formula
known to us for the $SU(3)$ integral. To obtain a closed form for the index of
the strongly-coupled CFTs appearing in higher rank theories one has
 to learn how to ``invert the superconformal tails''.

The superconformal index of the generalized quiver theories can be
built from  a small number of building blocks, the structure
constants and the metric of section~\ref{sdualSec}. We have
explicitly shown, at least in perturbation theory in the chemical
potential $t$, that the superconformal index of these theories is
consistent with S-duality. These structure constants
and metric can be interpreted as defining a $2d$ topological quantum
field theory, generalizing to $A_2$ the construction given in \cite{Gadde:2009kb} for $A_1$.
 It would be very interesting to obtain a Lagrangian
description for these TQFTs, perhaps by direct dimensional
reduction of the twisted $(2,0)$ theory on $S^3 \times S^1$.

Finally, from a pure mathematics viewpoint, we have seen that
 S-duality implies a number of identities that must be obeyed by integrals of elliptic Gamma functions
and that we have checked perturbatively. We 
 collect the exact  identities   in appendix~\ref{identApp}. It would
 be nice to find analytic proofs.

\section*{Acknowledgements}
We thank Davide Gaiotto and Yuji Tachikawa for useful discussions.
This work was  supported in part by DOE grant DEFG-0292-ER40697 and by NSF grant PHY-0653351-001. Any
opinions, findings, and  conclusions or recommendations expressed in this
material are those of the authors and do not necessarily reflect the views of the National
Science Foundation.

\appendix
\section{$t$ expansion in the weakly-coupled frame}\label{LHS}
We expand the index~\eqref{indexWeak} in $t$
as
\be
\mathcal{I}_{a,{\mathbf z};b,{\mathbf y}}=\sum_{k=0}^\infty b_k\, t^k.
\ee
The first few orders are
\be
      \label{eq:expansion:LHS:full}
        &&b_0=1,\nonumber\\
        &&b_1=b_2=b_3=0,\nonumber\\
        &&b_4
         =\frac{1}{v}\chi^{SU(6)}_{\mathbf{35},adj}+\frac{1}{v}+v^2,\nonumber\\
        &&b_5=-v\left(y+\frac{1}{y}\right),\\
        &&b_6
         =\frac{1}{v^{3/2}}\chi^{SU(6)}_{\mathbf{20}}\left(\left(\frac{a}{b}\right)^{3/2}+
\left(\frac{b}{a}\right)^{3/2}\right)
          -\chi^{SU(6)}_{\mathbf{35},adj}+v^3-1,\nonumber\\
        &&b_7
         =\frac{1}{v}\left(y+\frac{1}{y}\right)\chi^{SU(6)}_{\mathbf{35},adj}
          +\frac{2}{v}\left(y+\frac{1}{y}\right),\nonumber\\
        &&b_8
         =\frac{1}{v^2}\chi^{SU(6)}_{sym^2\mathbf{35}}+v\chi^{SU(6)}_{\mathbf{35},adj}
          -\frac{1}{\sqrt{v}}\chi^{SU(6)}_{\mathbf{20}}\left(\left(\frac{a}{b}\right)^{3/2}+
\left(\frac{b}{a}\right)^{3/2}\right)
          +v^4-v\left(y+\frac{1}{y}\right)^2+2v,\nonumber\\
        &&b_9
         =-2\left(y+\frac{1}{y}\right)\chi^{SU(6)}_{\mathbf{35},adj}
          +\frac{1}{v^{3/2}}\left(y+\frac{1}{y}\right)\chi^{SU(6)}_{\mathbf{20}}\left(\left(\frac{a}{b}\right)^{3/2}+
\left(\frac{b}{a}\right)^{3/2}\right)
          -2\left(y+\frac{1}{y}\right).\nonumber
\ee
 In the above equation we decomposed $SU(6)\supset SU(3)_z\otimes
  SU(3)_{y^{-1}}\otimes U(1)$. The branching of $\mathbf{35}$ and $\mathbf{20}$
  of $SU(6)$ is given by (see \cite{Slansky:1981yr}),
\be
\mathbf{35}&=&(\mathbf{1},\mathbf{1})_0+(\mathbf{8},\mathbf{1})_0+(\mathbf{1},\mathbf{8})_0
                +(\mathbf{\bar 3},\mathbf{{3}})_2+(\mathbf{{3}},\mathbf{\bar 3})_{-2}\,,\\
\mathbf{20}&=&(\mathbf{1},\mathbf{1})_3+(\mathbf{1},\mathbf{1})_{-3}
                +(\mathbf{\bar3},\mathbf{3})_{-1}+(\mathbf{3},\mathbf{\bar 3})_1\,.\nonumber
\ee
  For example, the character of the adjoint is
\be
\chi^{SU(6)}_{\mathbf{35},adj}&=&\left[
          \left(a\,b\right)^{1/2}(z_1+z_2+z_3)+\left(a\,b\right)^{-1/2}(\frac{1}{y_1}+\frac{1}{y_2}+\frac{1}{y_3})
        \right]\times\,\\
        &&\qquad\qquad\times\left[
          \left(a\,b\right)^{-1/2}\left(\frac{1}{z_1}+\frac{1}{z_2}+\frac{1}{z_3}\right)
          +\left(a\,b\right)^{1/2}\left(y_1+y_2+y_3\right)
        \right]-1\,.\nonumber
\ee We conclude that the $U(1)$ charge in $SU(6)$ can be
  identified as $(a\,b)^{-1/2}$.

\section{Inversion theorem}\label{inversionSec}

\begin{figure}[htbp]
\begin{center}
\epsfig{file=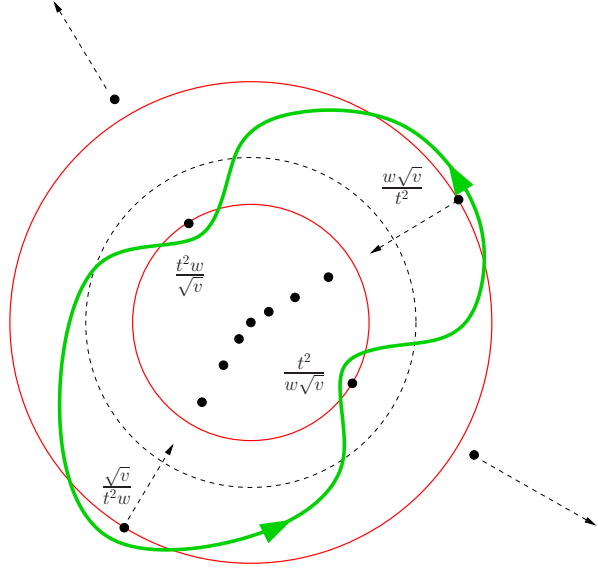,scale=0.5}
\end{center}
 \begin{center}
\caption{The integration contour $C_w$ (green). The dashed (black) circle is the unit circle ${\mathbb T}$.
Black dots are poles of $\Gamma\left(\frac{\sqrt{v}}{t^2}\,w^{\pm1}\,z^{\pm 1}\right)$.
 There are four sequences of poles:
two sequences starting at $\frac{\sqrt{v}}{t^2}w^{\pm1}$ and converging to $z=0$, and
two sequences starting at $\frac{t^2}{\sqrt{v}}w^{\pm1}$ and converging to $z=\infty$.
 The contour encloses the two former sequences.
} \label{contour}
\end{center}
\end{figure}

In this appendix we quote the inversion theorem~\cite{spirinv}, which we use in section~\ref{strongSec}
to obtain the index of the $E_6$ theory.
Define
\be
\delta(z,\,w; T) \equiv \frac{\Gamma(T\, z^{\pm1}\,w^{\pm1};\,p,q)}{\Gamma(T^2,\,z^{\pm2};\,p,q)}.
\ee
 If $T$, $p$ and $q$ are such that
\be |\text{max}(p,\,q)| < |T|<1\,, \ee then the following theorem
holds true. For fixed $w$ on the unit circle we define a contour
$C_w$ (see figure~\ref{contour}) in the annulus ${\mathbb
A}=\{|T|-\epsilon<|z|<|T|^{-1}+\epsilon\}$ with small but finite
$\e\in {\mathbb R}^+$, such that the points $T^{-1}w^{\pm1}$ are in
its interior and $C_w=C^{-1}_w$ (i.e. an inverse of the point in the
interior of $C_w$ is in the exterior of $C_w$). Let $f(z)=f(z^{-1})$
be a holomorphic function in ${\mathbb A}$. Then for
$|T|<|x|<|T|^{-1}$, \be\label{inv} \hat f(w)=\kappa\oint_{C_w}\frac{dz}{2\pi
i\,z}\,\delta(z,w;, T^{-1})\, f(z)\quad \longrightarrow\quad
f(x)=\kappa\oint_{{\mathbb T}}\frac{dw}{2\pi i\,w}\,\delta(w,x;, T)\, \hat f(w)\,.\nonumber\\
\ee

Our expression for the index in the strongly-coupled frame~\eqref{strong} is of the form of the right hand side of~\eqref{inv}.
Thus, to use the inversion theorem to obtain the index of $E_6$ theory we \textit{assume} that this index can be written 
as 
\be
\Gamma(t^{2}v\,w^{\pm2})\;C^{(E_6)}\left((w,\,r),\mathbf{y},\mathbf{z}\right)=\kappa\oint_{C_w}\,\frac{d s}{2\pi
i\,s}\,\frac{\Gamma(\frac{\sqrt{v}}{t^2}w^{\pm1}\,s^{\pm1})}{\Gamma(\frac{v}{t^4},\;s^{\pm2})}\,
F\left(s,r;\mathbf{y},\mathbf{z}\right)\,,
\ee for some function $F$. The theorem~\eqref{inv} then implies that $F\left(s,r;\mathbf{y},\mathbf{z}\right)=\hat{\mathcal I}\left(s,r;\mathbf{y},\mathbf{z}\right)$ with $\mathcal{I}\left(s,r;\mathbf{y},\mathbf{z}\right)$ given in~\eqref{strong}.

\section{The Coulomb and Higgs branch operators of $E_6$
SCFT}\label{CoulHiggs}

We collect here a few facts about the Coulomb and the Higgs branches of $E_6$ SCFT, following the analysis of~\cite{Gaiotto:2008nz}.
 Argyres-Seiberg duality can be used to determine the quantum
numbers of protected operators of $E_6$ theory if their dual
operators in the dual $SU(3)$ theory are known. The Coulomb branch operator
$u$ of the $E_6$ theory (the operator whose vev parametrized the Coulomb branch)
is identified as $\tr\, \phi^3$ in the
$SU(3)$ theory. Since $\phi$ has quantum numbers
$(E,j_1,j_2,R,r)=(1,0,0,0,-1)$, $u$ should have quantum numbers
$(3,0,0,0,-3)$ and contribute to the superconformal index as $t^6v^3$.

The  operator $\mathbb{X}$ whose vev parametrized the Higgs branch transforms in the adjoint
representation of $E_6$. Under the $SU(2)\otimes
SU(6)$ subgroup of $E_6$ it decomposes as
\begin{equation}
  X^i_j,\,\,\,\,\,Y^{[ijk]}_\alpha,\,\,\,\,\,Z_{\alpha\beta}\,,
\end{equation}
where $i,j,k=1,\ldots,6$ are the $SU(6)$ indices, and $\alpha,\beta=1,2$
are the $SU(2)$ indices. At the same time, the $SU(2)$ gauge theory provides the
quarks $q_\alpha$, $\tilde{q}_\alpha$ and the $F$-term constraint
\begin{equation}
  Z_{\alpha\beta}+q_{(\alpha}\tilde{q}_{\beta)}=0\,.
\end{equation}
Thus the gauge-invariant operators are
\begin{equation}
   (q\tilde{q}),\,\,\,\,\,X^i_j,\,\,\,\,\,(Y^{ijk}q),\,\,\,\,\,(Y_{ijk}\tilde{q})\,.
\end{equation}

On the $SU(3)$ side, the Higgs branch is parameterized by gauge
invariant operators
\begin{equation}
  M^i_j=Q^i_a\tilde{Q}^a_j,\,\,\,\,\,B^{ijk}=\epsilon^{abc}{Q}_a^i{Q}_b^j{Q}_c^k,\,\,\,\,\,
  \tilde{B}_{ijk}=\epsilon_{abc}\tilde{Q}^a_i\tilde{Q}^b_j\tilde{Q}^c_k\,,
\end{equation}
where $Q^i_a$ and $\tilde{Q}^a_i$ are the squark fields,
$i=1,\ldots,6$ are flavor indices, and $a=1,2,3$ the color indices.

The duality of the two sides suggests the following identification
\begin{eqnarray}
  \tr
  M\leftrightarrow(q\tilde{q}),&\,\,\,\,\,&\hat{M}^i_j\leftrightarrow
  X^i_j,\\
  B^{ijk}\leftrightarrow(Y^{ijk}q),&\,\,\,\,\,&\tilde{B}_{ijk}\leftrightarrow(Y_{ijk}\tilde{q})
\end{eqnarray}
where $\hat{M}^i_j$ is the traceless part of $M^i_j$. Since the
quantum numbers of $Q$ are $(1,0,0,1/2,0)$, the quantum numbers of
$\mathbb{X}$ should be $(2,0,0,1,0)$, and contribute to the  index as $t^4/v$.

\section{Identities from S-duality}\label{identApp}

In this appendix we summarize identities of integrals of elliptic Gamma functions implied
by S-duality of the $SU(3)$ quiver theories.
\subsection*{Generalization of~\cite{Focco}}
We define
\begin{eqnarray}\label{deBultSUn}
&&\mathcal{I}^{(n)}\left(a\,,{\mathbf z}_{SU(n)};b\,,{\mathbf y}_{SU(n)}\right) \equiv \frac{2^{n-1}}{n!}\kappa^{n-1}\Gamma(t^{2}v)^{n-1}\times\\
&&
 \left.\oint_{\mathbb{T}^{n-1}}\prod_{i=1}^{n-1}\frac{dx_{i}}{2\pi i\,x_{i}}
\frac{\prod_{i=1}^n\prod_{j=1}^n\Gamma\left(\frac{t^{2}}{\sqrt{v}}\left(\frac{az_{i}}{x_{j}}\right)^{\pm1}\right)
\Gamma\left(\frac{t^{2}}{\sqrt{v}}\left(b\,y_{i}\,x_{j}\right)^{\pm1}\right)\prod_{i\neq j}
\Gamma\left(t^{2}v\,\frac{x_{i}}{x_{j}}\right)}{\prod_{i\neq j}\Gamma\left(\frac{x_{i}}{x_{j}}\right)}\right|_{\prod_{j=1}^nx_j=1}.\nonumber
\end{eqnarray}
The claim is that
\be
\mathcal{I}^{(n)}\left(a\,,{\mathbf z}_{SU(n)};b\,,{\mathbf y}_{SU(n)}\right)=
\mathcal{I}^{(n)}\left(b\,,{\mathbf z}_{SU(n)};a\,,{\mathbf y}_{SU(n)}\right)\,.
\ee For $SU(2)$ this identity was proven in~\cite{Focco}, and for $SU(3)$ we have performed perturbative checks.
The usual S-duality of $N_f=2n$ $SU(n)$ theories implies that this identity should be true for any $n$.
Note that for $t=v$ this is a special case of  identities discussed in~\cite{rains}.

\subsection*{$E_6$ Integral}
We define
\begin{equation}
\begin{gathered}
 C^{(E_6)}\left((w,\,r),\mathbf{y},\mathbf{z}\right)\equiv \frac{2\,\kappa^3\Gamma(t^{2}v)^{2}}{3\,\Gamma(t^{2}v\,w^{\pm2})}\;
\oint_{C_w}\,\frac{d s}{2\pi i\, s}\,\frac{\Gamma(\frac{\sqrt{v}}{t^2}w^{\pm1}\,s^{\pm1})}{\Gamma(\frac{v}{t^4},\;s^{\pm2})}\,\times
\\
\times\;\oint_{\mathbb{T}^{2}}\prod_{i=1}^{2}\frac{dx_{i}}{2\pi i \,x_{i}}
\frac{\displaystyle\prod_{i=1}^3\prod_{j=1}^3\Gamma\left(\frac{t^{2}}{\sqrt{v}}\left(\frac{s^{\frac{1}{3}}\,z_{i}}{x_{j}\,r}\right)^{\pm1}\right)
\Gamma\left(\frac{t^{2}}{\sqrt{v}}\left(\frac{s^{-\frac{1}{3}}\,y_{i}\,x_{j}}{r}\right)^{\pm1}\right)
\prod_{i\neq j}\Gamma\left(t^{2}v\frac{x_{i}}{x_{j}}\right)}{\displaystyle\prod_{i\neq j}\Gamma\left(\frac{x_{i}}{x_{j}}\right)}\,.
\end{gathered}
\end{equation}  This integral
has manifest symmetry under  $SU(2)_w \otimes SU(6)$, where the $SU(6)$ has been decomposed as
 $SU(3)_{\mathbf{z}}\otimes SU(3)_{\mathbf{y}^{-1}}\otimes U(1)_r$.
The identification with the index of the $E_6$ SCFT implies that there
 must be a symmetry enhancement $SU(2)_w \otimes SU(6) \to E_6$.
Two properties that are sufficient to guarantee $E_6$ covariance are: first, 
\be
&&C^{(E_6)}\left((w,\,r),\mathbf{y},\mathbf{z}\right) =C^{(E_6)}\left(\left(\frac{w^{1/2}}{r^{3/2}},\,\frac{1}{w^{1/2}\,r^{1/2}}\right),\mathbf{y},\mathbf{z}\right)\, ,
\ee 
which is the statement that $(w,r)$ combine into a character of $SU(3)$ (which we shall denote by $\mathbf{w}$);
second,
\be
C^{(E_6)}(\mathbf{w},\mathbf{y},\mathbf{z})=C^{(E_6)}(\mathbf{y},\mathbf{w},\mathbf{z})\,.
\ee
We presented perturbative evidence for the full $E_6$ symmetry in the text.

\subsection*{S-dualities of $SU(3)$ quivers}
Define
\be{\mathcal I}_{3333}\left(\mathbf{y},\mathbf{z},\mathbf{u},\mathbf{s}\right) &\equiv&
\oint_{\mathbb{T}^2}\prod_{i=1}^2\frac{dx_i}{2\pi i x_i}\prod_{i\neq j}
\frac{\Gamma\left(t^2v x_i/x_j\right)}{\Gamma\left( x_i/x_j\right)}
C^{(E_6)}\left(\mathbf{y},\mathbf{z},\mathbf{x}\right)
C^{(E_6)}\left(\mathbf{u},\mathbf{s},\mathbf{x}^{-1}\right)\,,\\
{\mathcal I}_{3331}\left(\mathbf{y},\mathbf{z},\mathbf{u},a\right) &\equiv&
\oint_{\mathbb{T}^2}\prod_{i=1}^2\frac{dx_i}{2\pi i x_i}\prod_{i\neq j}
\frac{\Gamma\left(t^2v x_i/x_j\right)}{\Gamma\left( x_i/x_j\right)}
C^{(E_6)}\left(\mathbf{y},\mathbf{z},\mathbf{x}\right)
\prod_{i,j=1}^3\Gamma\left(\frac{t^2}{\sqrt{v}}\left(a\, x_i^{-1}\,u_j\right)^\pm\right)\,.\nonumber
\ee
The S-dualities of the $SU(3)$ quivers imply
\be
&&{\mathcal I}_{3333}\left(\mathbf{y},\mathbf{z},\mathbf{u},\mathbf{s}\right)=
{\mathcal I}_{3333}\left(\mathbf{y},\mathbf{u},\mathbf{z},\mathbf{s}\right)\,,\\
&&{\mathcal I}_{3331}\left(\mathbf{y},\mathbf{z},\mathbf{u},a\right)=
{\mathcal I}_{3331}\left(\mathbf{y},\mathbf{u},\mathbf{z},a\right)\,.\nonumber
\ee

\bibliography{sduality}
%\bibliography{ref}
\bibliographystyle{JHEP}
%\bibliographystyle{unsrt}
%\bibliography{ref}
%\bibliographystyle{apsrev}
%\bibliographystyle{plain}
%\bibliographystyle{utphys}

\end{document}